\documentclass[longbib,modern]{aastex701}
\journalinfo{}
\makeatletter
\def\frontmatter@title@above{}%
\makeatother

\usepackage{natbib}
\usepackage[version=4]{mhchem} % Formula subscripts using \ce{}
\usepackage{graphicx}	% Including figure files
\usepackage{amsmath}	% Advanced maths commands
\usepackage{amssymb}	% Extra maths symbols
\usepackage{lipsum}
\usepackage{subfig}
\usepackage{xcolor}
\usepackage{orcidlink} %% orcid

\renewcommand{\thefootnote}{\alph{footnote}}

\makeatletter
\renewcommand{\doi}[1]{\@ifnextchar.{\@gobble}{}}
\renewcommand{\url}[1]{\@ifnextchar.{\@gobble}{}}
\makeatother

\begin{document}

\author[0000-0003-4224-6829]{Brandt A. L. Gaches}
\affiliation
{Faculty of Physics, University of Duisburg-Essen, Lotharstraße 1, 47057 Duisburg, Germany}
\email[show]{brandt.gaches@uni-due.de}
\author[0000-0001-8504-8844]{Serena Viti}
\email{}
\affiliation{Leiden Observatory, Leiden University, P.O. Box 9513, 2300 RA Leiden, The Netherlands}
\affiliation{Transdisciplinary Research Area (TRA) ‘Matter’/Argelander-Institut für Astronomie, University of Bonn, Auf dem H{\"u}gel 71, 53121, Bonn, Germany}
\affiliation{Department of Physics and Astronomy, University College London, Gower Street, London WC1E 6BT, UK}

\correspondingauthor{Brandt A. L. Gaches}

\title{High-energy astrochemistry in the molecular interstellar medium}

\begin{abstract}
In the past decade, there has been a significant shift in astrochemistry with a renewed focus on the role of non-thermal processes on the molecular interstellar medium, in particular energetic particles (such as cosmic ray particles and fast electrons) and X-ray radiation. This has been brought about in large part due to new observations of interstellar complex organic molecules (iCOMS) in environments that would inhibit their formation, such as cold, dense gas in prestellar cores or in the highly energetic environments in galactic centers. In parallel, there has been a plethora of new laboratory investigations on the role of high-energy radiation and electrons on the chemistry of astrophysical ices, demonstrating the ability of this radiation to induce complex chemistry. In recent years, theoretical models have also begun to include newer cosmic-ray-driven processes in both the gas and ice phases. In this review, we unify aspects of the chemistry driven by X-ray radiation and energetic particles into a ``high-energy astrochemistry'', defining this term and reviewing the underlying chemical processes. We conclude by examining various laboratories where high-energy astrochemistry is at play and identify future issues to be tackled.

\vspace{0.25cm}
\includegraphics[width=\textwidth]{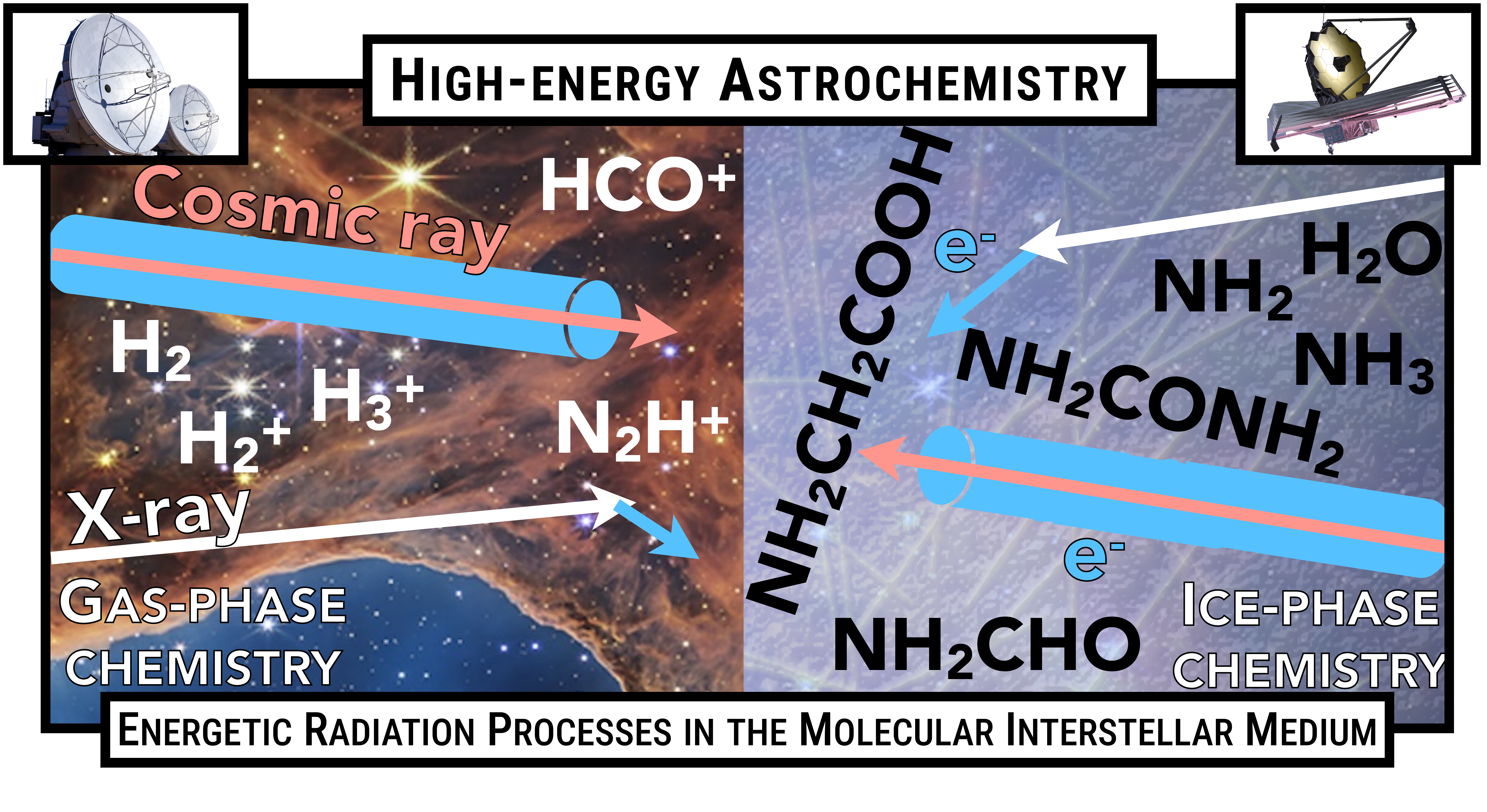}
\end{abstract}


\section{Introduction}
The molecular interstellar medium is impacted by a wide range of physical and radiative processes. In molecular clouds, condensations in the interstellar medium in which hydrogen is primarily in molecular form, \ce{H2}, the gas is cold (10 Kelvin or lower) and relatively dense (number densities exceeding 100 cm$^{-3}$). Due to these cold temperatures, the gas-phase chemistry occurs primarily through ion-neutral reactions \citep{Tielens2013}, which are usually barrier-free. As such, in regions shielded from ultraviolet radiation sources, high-energy radiation drives the chemistry as the primary ionization source. The chemistry of the molecular interstellar medium is diverse and complex, ranging from diatomic molecules -- including hydrides with heavy atoms -- to interstellar complex organic molecules (iCOMS) -- organic molecules containing six or more atoms \citep{Herbst2009} -- to complex hydrocarbons and carbon chains. There is also a vast range of environments in which chemistry occurs, enabling many interesting laboratories of key chemical and physical processes. Many of these environments sit in regions of galaxies immersed in high-energy radiation, from galactic cosmic rays throughout the interstellar medium, to in energetic environments such as near active galactic nuclei or high-mass star-forming regions. One thing is quite clear: the high-energy universe often impinges on the molecular universe.

\subsection{Sources of high-energy radiation}
There are a variety of sources of X-ray radiation and particle acceleration that are important for the interstellar medium: \\
{\textbf{Active galactic nuclei:}} It is likely that the centers of most galactic bulges host supermassive black holes \citep{Kormendy2013, Heckman2014}. When these supermassive black holes actively accrete, they can become active galactic nuclei (AGN), which become intensely powerful radiation sources across the electromagnetic spectrum. AGN are bright hard X-ray sources as a result of Comptonization of photons originating from the accretion disk by hot electrons in the surrounding corona \citep{Bianchi2022}. The X-rays from AGN can dominate the ionization of the gas in the inner nucleus since hard X-rays can propagate much further than the accretion-powered UV. Their X-ray spectrum is characterized by a power-law in hard X-rays, which can be attenuated by gas within the inner few pc. AGN have also been found to accelerate cosmic rays up to high energies \citep{Eichmann2022, Rieger2022}, becoming important sources of cosmic-ray ionization in their central nucleus \citep{Papadopoulos2010, Gonzalez-Alfonso2013, Koutsoumpou2025}. There are a number of local AGN that have been intensely studied: the Seyfert-2 AGNs in the composite galaxy NGC 1068 and in Circinus are, for example, great astrochemical laboratories to study the effects of X-rays and cosmic rays on the surrounding gas. Ultra-Luminous Infrared Galaxies (ULIRG), such as Mrk231 and Arp220, also host powerful AGNs, although objects such as Arp220 are heavily dust-obscured and hence characterizing the energy output from their AGN is harder.    \\
{\textbf{Supernova remnants:}} Supernova remnants have been known to be important sources of both X-rays and cosmic rays. Shocks propagating through the surrounding ISM accelerate particles to high energies. While it has been historically thought that supernovae were the dominant source of galactic cosmic rays \citep{Blasi2013}, this has recently come into debate, especially for the lower energy sub-GeV cosmic rays important for chemistry \citep{Gabici2019}. Hot gas within the cavity becomes X-ray bright through both thermal and non-thermal processes \citep{Vink2012, Vink2022, Yamaguchi2022}.\\
{\textbf{High-mass stars:}} High-mass stars emit powerful winds into their surrounding region. X-ray emission occurs throughout their lifespan \citep{Rauw2022}, including evolved stages such as Wolf-Rayet stars. Magnetized high-mass stars can produce variable X-ray emission through magnetic reconnection. Furthermore, massive stars in binaries can become important sources of X-ray radiation when their winds accrete onto compact objects such as neutron stars, powering intense high-mass X-ray binaries \citep{Fornasini2023}, which radiate with a hard X-ray power-law spectrum. The HII regions around high-mass stars have recently been proposed to be an acceleration site for cosmic rays \citep{Meng2019, Padovani2019}. Finally, the collective winds from massive stellar clusters have been proposed as an important site of cosmic ray acceleration for decades \citep{Casse1980, Parizot2004, Aharonian2019, Gabici2023}.\\
{\textbf{Protostars: }} Protostars have been known to produce X-ray radiation for decades, powered by a combination of accretion and magnetic reconnection \citep{Feigelson1999, Feigelson2007, Hartmann2016, Getman2021}. In fact, surveys of star-forming regions have demonstrated that these regions are brightly lit up in X-rays \citep{Townsley2014, Townsley2019, Feigelson2013}. Magnetic reconnection in young stellar objects has been proposed as an important source of energetic particles \citep{Lee1998}. Accretion-powered X-ray radiation exhibits a thermal spectrum while reconnection-powered radiation exhibits a power-law spectrum. More recently, actively accreting protostars were proposed as cosmic ray sources \citep{Padovani2015}, with acceleration occurring in their surface accretion shocks \citep{Padovani2016, Gaches2018} and emitted jets \citep{Padovani2016, Rodriguez-Kamenetzky2017, Osorio2017, Sanna2019}. Accretion-powered particle acceleration has not been directly confirmed by observations, although there have been indications of an enhanced cosmic-ray ionization rate towards embedded star formation \citep{Cabedo2023, Pineda2024}. 

\subsection{Observations of high-energy chemistry}

In recent years, a number of important observations of complex organic molecules in objects thought to be pristine, such as the comet 67P/Churyumov-Gerasimenko \citep{Altwegg2016} and cold, dense prestellar cores \citep{Scibelli2020}, have led to a new foundational model in astrochemistry which now includes a variety of non-thermal processes. Much of the advancements have been in ice chemistry, driven in large part by observational and laboratory studies, with further developments in the gas phase, including more sophisticated radiation transport processes. 

In the gas phase, high-energy chemistry is characterized by its role in driving ion-neutral chemistry through regulating the ionization fraction. These ion-neutral reactions are important since they proceed largely without an energy barrier at the Langevin rate ($\approx 10^{-9}$ cm$^{3}$ s$^{-1}$). Beyond ionizing, cosmic rays can also dissociate molecular bonds, in particular through secondary electron impact dissociation. \ce{He+} is also an important molecular bond destroyer, and in molecular clouds is only produced appreciably by high-energy radiation due to the high ionization potential of helium (24.59 eV). Doubly ionized species can also be produced from X-ray ionization through the Auger process.

Observationally, \ce{H3+} and simple light hydride ions such as \ce{OH+}, \ce{H2O+}, \ce{H3O+} and \ce{ArH+} can be used to directly trace the ionization fraction and ionization rate. These ions are observed through absorption lines \citep{Indriolo2012, Gonzalez-Alfonso2013, Schilke2014, Indriolo2015}, and so primarily constrain the ionization rate in diffuse envelopes around molecular clouds. There has been debate on what the average ISM cosmic-ray ionization rate is in the Milky Way, with some observational evidence pointing more to a Voyager-like ionization rate, producing an ionization rate $\zeta \approx 10^{-17}$, or a more elevated rate seen through absorption studies along sight lines \citep{Indriolo2012, Indriolo2015, Neufeld2017}. Many of these absorption line studies utilized density estimates from \ce{C2} observations. Recent re-analysis of these data with new \ce{C2} collisional rates has brought down the diffuse gas measurements by factors of a few \citep{Neufeld2024}. Recently, the combination of 3D dust maps of the local interstellar medium and Gaia catalogs has enabled unique and high fidelity measurements of the cosmic-ray ionization rate through H3+ in the local interstellar medium \citep{Obolentseva2024, Indriolo2025}.

These measurements, in principle, measure the total ionization rate, indiscriminately of whether it originates from X-rays or cosmic rays. However, the lack of X-ray sources nearby suggests that these ionization rates are most likely due to cosmic rays. X-ray ionization may dominate in regions close to  galactic centers, such as NGC 1068 \citep{Viti2014}. When cosmic rays appreciably heat the gas, it can also impact molecular line ratios such as \ce{HNC/HCN}, due to the temperature dependence of the conversion between these isomers\citep{Behrens2024}.

In the solid phase, the primary driver of chemistry is the induced electron cascade \citep{Mason2014, Oberg2016, Dartois2019}. Recent surveys of cold, dense cores have shown an abundance of iCOMS \citep{Scibelli2020, Jimenez-Serra2021, Scibelli2024}. Recently, volatile glycine (\ce{NH2CH2COOH}) was detected in the comet 67P/Churyumov-Gerasimenko along with \ce{CH3NH2} and \ce{CH3CH2NH2} \citep{Altwegg2016}. The volatiles are thought to come from pristine gas.  Non-thermal processes in the ices, driven through the secondary electron cascade following direct ionization of the ice and the mineral dust grain, can provide the energy to build complex organic molecules \citep{Oberg2016, Arumainayagam2019}. Cosmic rays, especially heavy ions, also induce sputtering \citep{Dartois2019, Dartois2020} and alter the ice structure \citep{Dartois2015}. Alongside high-energy processes, there have also been recent developments with the newly proposed non-diffusive processes \citep{Garrod2022, Maitrey2025} and freeze-out of atomic species such as carbon \citep{Ferrero2024}, which have also been shown to build complexity in cold, unprocessed ices. 

\subsection{Differences in X-ray and cosmic-ray transport}
One of the main fundamental differences in how X-rays and cosmic rays interact with gas is how they are transported. In molecular clouds, X-rays are absorbed by the intervening gas and potentially scattered at high energies. In the regime of no scattering, the X-ray flux, $F_X(E)$ (erg s$^{-1}$ cm$^{-2}$ eV$^{-1}$), from a point source can be simply described by its luminosity, $L_X(E)$ (erg s$^{-1}$ eV$^{-1}$), the distance from the source, $d$, and the opacity of the intervening gas, $\tau_X$, e.g., $F(E) \approx \frac{L(E)}{4\pi d^2} e^{-\tau_X}$. For slab irradiation, it is even simpler, where in the regime of no scattering, $F(E) = F_0(E) e^{-\tau_X}$, for some surface irradiating flux, $F_0$.

Cosmic rays are tightly coupled to the magnetic fields and can undergo diffusion (scattering on magnetic fields), (re)acceleration, and attenuation by interactions with matter and radiation fields \citep{Schlickeiser2002, Longair2011}. The transport of cosmic rays is far from a trivial problem, but under the assumption that only continuous losses are dominant \citep{Padovani2009}, the particle flux spectrum, $j(E)$ (particles cm$^{-2}$ s$^{-1}$ eV${-1}$ sr$^{-1}$), can be solved simply with knowledge of the energy loss function. However, magnetic field effects can cause complex transport regimes, potentially excluding cosmic rays or changing the diffusion coefficient. X-rays are not impacted by this, and so there can be environments in which cosmic rays cannot penetrate a molecular cloud, but X-rays can.

\subsection{Unifying high-energy astrochemistry}

\begin{figure}
    \centering
    \includegraphics[width=\columnwidth]{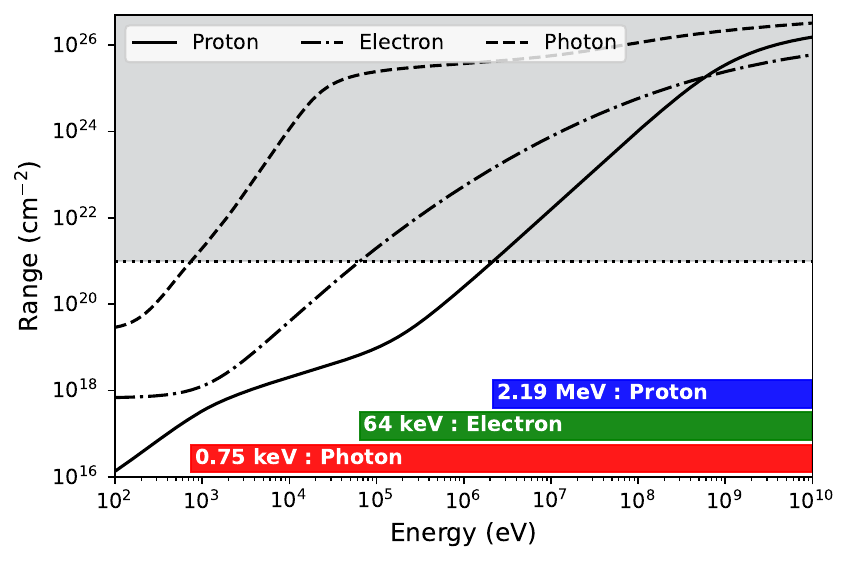}
    \caption{\label{fig:range}The stopping column, or range, in units of cm$^{-2}$ as a function of energy, $E$, for protons, electrons, and photons moving through cold molecular gas. The gray shaded region denotes $N(H) > 10^{21}$ cm$^{-2}$. The colored boxes denote the minimum energy for that species such that the range exceeds $10^{21}$ cm$^{-2}$. Range functions are from \citet{Padovani2018}.}
\end{figure}

There have been a number of key reviews of photochemistry \citep{Oberg2016, Arumainayagam2019, Wolfire2022} and particle-driven chemistry \citep{Dalgarno2006, Indriolo2013, Padovani2020, Padovani2024, Arumainayagam2019}. Typically, X-ray radiation chemistry is treated as an extreme version of the UV chemistry observed in photo-dissociation regions (PDRs), although discussions of ice chemistry have started to unify X-ray and particle chemistry \citep{Vasconcelos2017, Arumainayagam2019}. In this review, we present the unifying concept of ``high-energy astrochemistry'', which we define as \emph{``the chemistry induced by ionizing radiation in which the primary interaction energy is sufficient to lead to a cascade of numerous secondary electrons ($N_e >> 1$)''}. The latter requirement excludes much of the UV and soft X-ray, where a single ionization or excitation event causes enough energy loss to inhibit further ionizations. This clarification effectively selects radiation chemistry in which there is a substantial (and sometimes dominating) contribution from secondary processes. Figure \ref{fig:range} compares the stopping columns, or ranges, in units of cm$^{-2}$ for cosmic ray protons and electrons, and photons, and compares them against a column density of $10^{21}$ cm$^{-2}$, relevant for transport in molecular clouds.

In this review, we start in Section \ref{sec:gasphase} detailing the underlying gas-phase chemical processes. We primarily focus on gas within the interstellar medium and molecular clouds, and refer the interested reader to other recent reviews in protoplanetary disk chemistry, such as \citet{Oberg2023}. In Section \ref{sec:icephase} we discuss ice processing by high-energy radiation. In Section \ref{sec:chemmodel}, we highlight the various ways in which these processes have been included in astrochemical models.  Finally, in Section \ref{sec:casestudies} we review different case studies of regions of interest in the molecular interstellar medium in which high-energy radiation can play dominant roles.

\section{Gas-phase processes}\label{sec:gasphase}
In this section, we describe the underlying chemical and thermodynamic processes brought on by high-energy radiation, highlighting in particular aspects which are similar and those which are distinct.
\subsection{Gas-phase Chemical processes}
Within the gas phase, the primary impact on the chemistry is through ionization of key abundant species. High-energy irradiation leads to the production of free electrons and induced FUV radiation. The former is crucial for determining the ionization fraction relevant for non-ideal hydrodynamic processes, while the latter is a crucial source of embedded FUV radiation in molecular clouds. These processes have been reviewed separately for cosmic rays \citep{Indriolo2013} and X-rays \citep{Wolfire2022}.

Cosmic ray particles, in particular sub-GeV protons and high-energy electrons, along with fast protons and heavy ions, directly ionize H, \ce{H2}, and He, with minor contributions due to direct ionization of other atoms and ionization and dissociation of abundant molecules such as \ce{CO}. Secondary electrons are also important for the ionization and dissociation of molecules, with models typically assuming that the ejected electron comes from the outermost orbital. The primary cosmic-ray \ce{H2} ionization rate, incorporating only primary protons, is
\begin{equation}
\zeta_p = A \times \int_{E_{\rm min}}^{E_{\rm max}} \sigma_p(E) j_p(E) dE,
\end{equation}
where $A$ is a constant relating to the geometry ($2\pi$ for slab irradiation and $4\pi$ for isotropic radiation), $\sigma_p(E)$ is the proton-impact \ce{H2} ionization cross section \citep{Rudd1988}, and $j_p(E)$ is the spectrum of primary protons. The non-relativistic proton-impact \ce{H2} ionization cross section takes the form \citep{Rudd1985b}
\begin{equation}
\sigma_p(E) = \frac{4\pi a_0^2}{\sigma_l^{-1} + \sigma_h^{-1}},
\end{equation}
where $\sigma_l(x) = C x^D$ and $\sigma_h(x) = [ A \ln(1 + x) + B]/x$, where $x = E/R$, $R = 13.6$ eV, $A = 0.71$, $B = 1.63$, $C = 0.51$, $D = 1.24$, and $a_0$ is the Bohr radius. 

\begin{figure}
    \centering
    \includegraphics[width=\columnwidth]{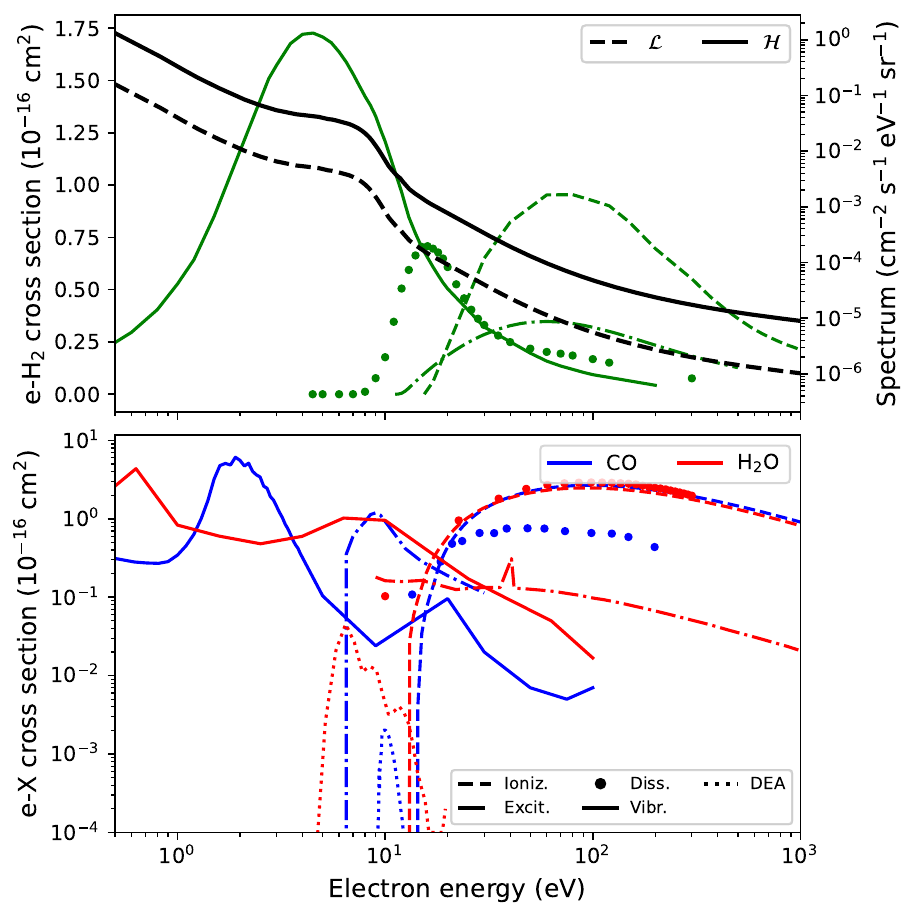}
    \caption{\label{fig:secelect}Top: Inelastic electron-impact cross sections for \ce{H2}, from \citet{Padovani2022} and \citet{Scarlett2023}. Secondary electron spectrum for the cosmic ray proton spectrum models \citep{Ivlev2015b, Padovani2022} $\mathcal{H}$ and $\mathcal{L}$ attenuated by a hydrogen nuclei column density of $N(H) = 10^{23}$ cm$^{-2}$ (black lines), from \citet{Padovani2022}. Bottom: Inelastic electron-impact cross sections for CO, from \citet{Itikawa2015}, and \ce{H2O}, from \citet{Song2021}.}
\end{figure}

The computation of secondary electrons is sensitive to the primary cosmic ray spectrum, and takes into account the differential electron-impact ionization cross sections of protons and electrons using electron-impact cross sections for ionization, dissociation, electronic excitation, and rovibrational excitation \citep{Ivlev2021, Padovani2022}. We briefly describe a simplified on-the-spot approximation \citep{Ivlev2015b} (see also \citet{Padovani2024} for review), where it is assumed all secondary electrons are produced and absorbed locally. The secondary electron spectrum, produced by primary protons, is then given by the balance of energy losses locally and production
\begin{equation}
    j_{\rm{sec, e}}(E_e) \approx \frac{E_e}{L_e(E_e)} \int_{I(\ce{H2})}^{\infty} j_p(E_p)\frac{\partial \sigma^{\rm ion}_{p,\ce{H2}}(E_p, E_e)}{\partial E_e} dE_p,
\end{equation}
where $E_e$ is the electron energy, $E_p$ is the proton energy, $j_p$ is the primary proton spectrum, $L_e(E_e)$ is the energy loss function for electrons for the gaseous composition, $\frac{\partial \sigma^{\rm ion}_{p,\ce{H2}}(E_p, E_e)}{\partial E_e}$ is the differential ionization cross-section, and $I(\ce{H2}) = 15.426$ eV is the \ce{H2} ionization potential. The above relationship is typically sufficient if only ionizing secondary electrons are of interest. However, for lower energies, this relationship breaks down, and one must consider a much more complete theory. In particular, electron-impact excitation of electronic and rovibrational excitations becomes important at low energies, and the treatment of catastrophic versus continuous losses \citep{Ivlev2021}. 
Figure \ref{fig:secelect} shows the secondary electron spectrum produced by cosmic ray protons after being attenuated by a hydrogen nuclei column density $N(H) = 10^{23}$ cm$^{-2}$, from \citet{Padovani2022}. The secondary electron spectrum was computed using the more complete methodology of \citet{Ivlev2021} with newer cross-section data for \ce{H2} interactions, presented in \citet{Padovani2022}. The secondary-electron-induced \ce{H2} ionization rate is then
\begin{equation}
    \zeta_{\rm sec,e} = 4\pi \int_{I(\ce{H2})}^{\infty} j_{\rm sec,e}(E_e) \sigma^{\rm ion}_{e, \ce{H2}}(E_e) dE,
\end{equation}
where $\sigma^{\rm ion}_{e, \ce{H2}}(E_e)$ is the total electron-impact ionization cross section of \ce{H2}. The factor of $4\pi$ is due to considering the secondary electrons as a locally isotropic source.

Cosmic ray protons and electrons can also excite other non-ionizing quantum transitions within molecules. Figure \ref{fig:secelect} also shows the \ce{H2} inelastic electron-impact cross sections for ionization, dissociation, excitation to the first excited electronic level, and an example rovibrational cross section. The bottom sub-panel shows inelastic cross sections for carbon monoxide \citep{Itikawa2015} and water \citep{Itikawa2005, Song2021}. Recently, \citet{Padovani2025} demonstrated that direct electronic excitation of \ce{H2} by low-energy cosmic ray protons may contribute as much to the total electronic excitation as secondary electrons.

For X-ray radiation, photo-ionization leads preferentially to ionizing K-shell electrons, resulting in a cascade of Auger ionizations. The resulting primary photo-electrons are energetic enough to cause a substantial number of secondary electrons. Secondary electrons are the most important means of ionization for H, \ce{H2}, He, and many molecules. The primary photoionization rate is
\begin{equation}
\zeta_i(E) = \int_{E_{\rm min}}^{E_{\rm max}} \sigma_{i, pe}(E) \frac{F(E)}{E} dE,
\end{equation}
where $\sigma_{i,pe}$ is the photoionization cross section and $F(E)$ is the X-ray flux. Photoionization cross sections and Auger yields are often taken from the work of \citet{Verner1995}. The secondary ionization rate for species $i$ per species $i$, with abundance $x_i$, is
\begin{equation}
\zeta_{i,sec} = \int_{E_{\rm min}}^{E_{\rm max}} \sigma_{pa}(E)F(E) \frac{E}{Wx_i} dE,
\end{equation}
where $\sigma_{pa}(E)$ is the photoelectric absorption cross section and $W$ is the mean energy per ion pair. $W$ has been computed and tabulated for mixtures of H-He and \ce{H2}-He in the canonical work of \citet{Dalgarno1999}. In principle, the full secondary-electron production theory of \citet{Ivlev2021} can be adapted to X-ray ionization by replacing the primary source term, as noted in their work, although at the time of this review, it has not been done yet. In many applications (e.g., \citet{Meijerink2005}), the \ce{H} ionization rate is calculated, and the ionization rate for species $i$ is computed by a ratio of the peak of their electron-impact ionization cross sections, $\sigma_{ei}$,
\begin{equation}
\zeta_i = \frac{\sigma_{ei,i}}{\sigma_{ei,H}}\zeta_H.
\end{equation}
There has been a substantial amount of work dedicated to computing electron-impact ionization cross sections, primarily using the binary-encounter Bethe model of \citet{Kim1994, Hwang1996}, with data compiled in public databases, including NIST\footnote{\url{https://physics.nist.gov/PhysRefData/Ionization/molTable.html}} and the recent Astrochemistry Low-energy electron Cross-Section (ALeCS) database\footnote{\url{https://github.com/AstroBrandt/ALeCS}} \citep{Gaches2024}. Due to Auger ionization, X-ray ionization can produce multiply ionized species, which quickly recombine to singly or doubly ionized species.

Regardless of the ionization source, \ce{H2} can become electronically excited, and then radiatively decay back to the ground state and emit FUV radiation in the Lyman-Werner bands \citep{Cravens1978}. The pioneering work of \citet{Prasad1983} laid the initial framework of cosmic-ray induced FUV radiation in shielded regions, now called the Prasad-Tarafdar effect, with later works providing further details that decade by \citet{Sternberg1987} and \citet{Gredel1987, Gredel1989}. Most recently, \citet{Heays2017} and \citet{Padovani2024b} have recalculated the induced FUV spectrum and photo-dissociation rates induced by cosmic rays, with the \citet{Heays2017} results compiled online\footnote{\url{https://home.strw.leidenuniv.nl/~ewine/photo/}} and the \citet{Padovani2024b} in their Appendix D. The most accurate calculation to date is by \citet{Padovani2024b}, thanks to newer more accurate cross section data \citep{Scarlett2023}, models of cosmic ray propagation \citep{Padovani2018, Padovani2022} and secondary-electron production \citep{Ivlev2021}. \citet{Padovani2024b} predicted the total integrated FUV flux as a function of column density and $R_V$,
\begin{equation}
    \Phi_{\rm UV}(N) = 10^3 \left ( c_0 + c_1 R_V + c_2 R_V^2 \right )\times \left ( \frac{\zeta(\ce{H2})(N)}{10^{-16}~{\rm s^{-1}}} \right ) ~{\rm photons \, cm^{-2} \, s^{-1}},
\end{equation}
with the fitted coefficients, $c_0 = 5.023$, $c_1 = -0.504$ and $c_2 = 0.115$, valid between $3.1 \leq R_V \leq 5.5$. Furthermore, low-energy induced electrons can excite rovibrational levels in \ce{H2} leading to characteristic NIR radiation \citep{Gredel1995, Bialy2020}. The NIR \ce{H2} lines have been demonstrated to be a potentially robust tracer of the cosmic-ray ionization rate \citep{Bialy2020, Padovani2022, Gaches2022b}. The CR-induced lines have now been directly and definitively detected in space using the James Webb Space Telescope in the nearby Bok globule Barnard 68 \citep{Bialy2025}. From \citet{Bialy2020}, the total NIR \ce{H2} brightness is directly proportional to the amount of gas and cosmic-ray ionization rate,
\begin{equation}
    I_{\rm tot, cr-e} = \frac{1}{4\pi} g N(\ce{H2})\psi\bar{E}\zeta(\ce{H2}),
\end{equation}
where $g$ accounts for dust extinction, $\psi \approx 5.8$ is the number of excitations per ionization, and $\bar{E} \approx 0.486$ eV is the mean transition energy. Electron rotational excitation of high-dipole moment molecules such as HCN has been shown to be potentially important in environments with higher densities and electron fractions \citep{Goldsmith2017}, although this methodology has not been well developed for non-thermal low-energy secondary electrons.

\begin{figure*}
    \centering
    \includegraphics[width=\textwidth]{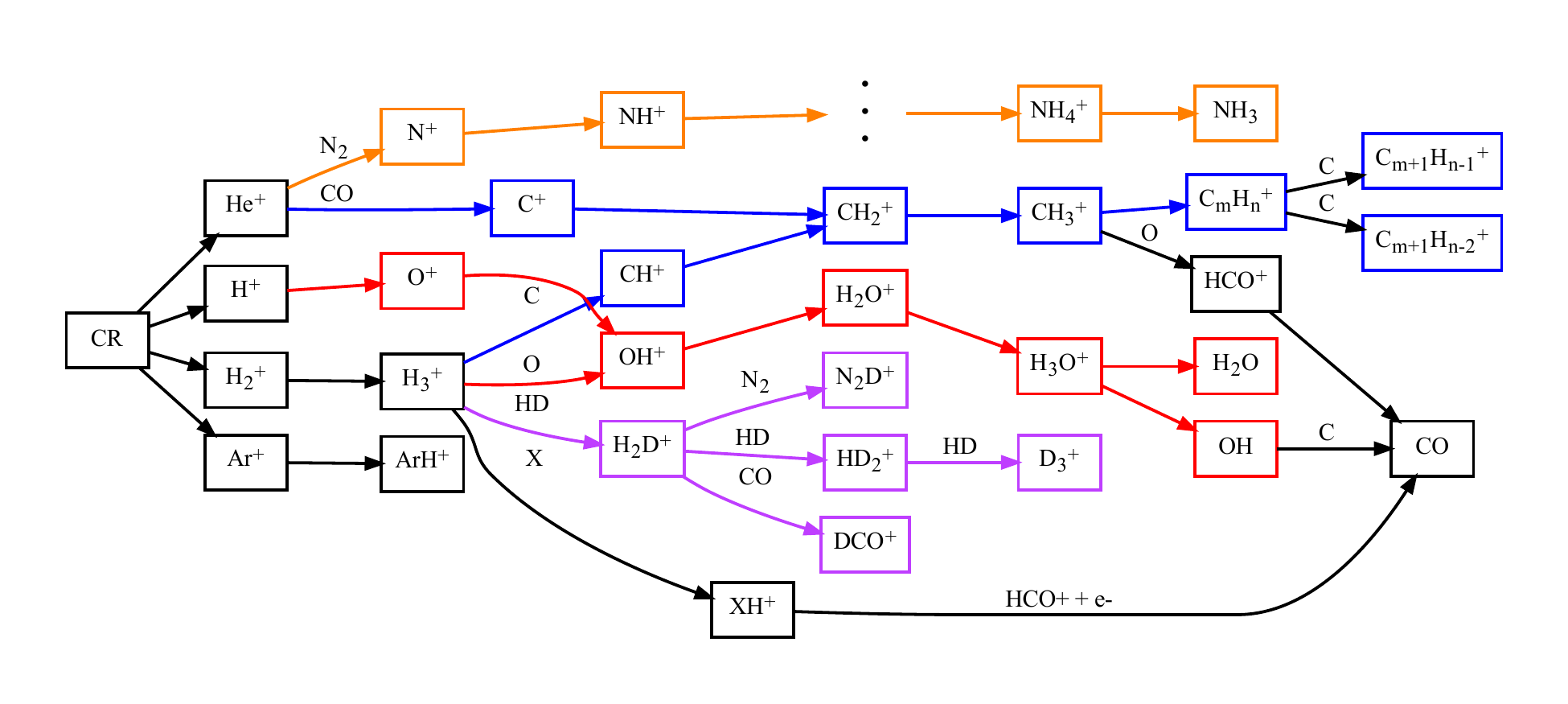}
    \caption{\label{fig:ionchemtree}Tree highlighting key ion-neutral pathways initiated by cosmic ray or X-ray ionization. Adapted from \citet{Padovani2024}.}
\end{figure*}
Following ionization, we summarize below the ion-neutral chemistry that occurs for important species. Figure \ref{fig:ionchemtree} highlights key pathways initiated by cosmic ray or X-ray ionization. One of the foundational reactions is with \ce{H2+},
\begin{equation}
\ce{H2+ + H2 -> H3+ + H}.
\end{equation}
This reaction is balanced by electron recombination,
\begin{equation}
\ce{H2+ + e- -> H + H},
\end{equation}
and the abundance of \ce{H3+} is likewise balanced by electron recombination
\begin{equation}
\ce{H3+ + e- -> H2 + H {\rm or} H + H +H}
\end{equation}
and proton-exchange reactions with neutral molecules
\begin{equation}
\ce{H3+ + X -> XH+ + H2}.
\end{equation}
Reactions with CO lead to \ce{HCO+} or \ce{HOC+}, with the production of \ce{HCO+} preferred over \ce{HOC+}, \ce{N2} leads to \ce{N2H+}, \ce{O} leads to \ce{OH+} or \ce{H2O+}, \ce{C} leads to \ce{CH+}, and Ar leads to \ce{ArH+}. Below, we detail the different branches of chemistry that can occur, but note that here we only summarize the initial beginnings of chemistry and do not endeavour to fully detail the ion-neutral chemical pathways.

Beyond constructive chemistry, it is worth highlighting the role of \ce{He+}, produced via ionization of neutral helium. \ce{He+} is highly efficient at destroying bonds, and in dense gas is one of the primary destroyers of CO via
\begin{equation}\label{eq:hepco}
\ce{He+ + CO -> He + O + C+}.
\end{equation}
This reaction also provides an important source of \ce{C+} in dense gas, where FUV photoionization of C is inefficient.

\subsubsection{Oxygen chemistry}
Following the ionization of neutral Oxygen, and the formation of \ce{OH+}, a set of hydrogenation reactions occur
\begin{equation}
\ce{OH+ + H2 -> H2O+ + H},
\end{equation}
\begin{equation}
\ce{H2O+ + H2 -> H3O+ + H}.
\end{equation}
These are in competition with dissociative recombination by electrons
\begin{equation}
\ce{H_nO+ + e- -> {\rm products}},
\end{equation}
where the products are, e.g., OH and \ce{H2O}. 

In diffuse or warm gas, the above chain can occur following the ionization of hydrogen,
\begin{equation}
\ce{H+ + O + \Delta E <=> O+ + H},
\end{equation}
\begin{equation}
\ce{O+ + H2 -> OH+ + H},
\end{equation}
where $\Delta E = 226$ K is the endothermicity of the forward reaction. Following the production of \ce{OH+}, the chemistry proceeds as explained above. This mechanism can be a source of \ce{H2O} in diffuse environments before the onset of freeze out, although it is worth noting that in dense gas, water primarily forms on grains. \ce{H2O} is also destroyed by \ce{He+}
\begin{equation}
\ce{He+ + H2O -> He + OH + H+ {\rm or} He + OH+ + H}.
\end{equation}
Finally, Reactions with \ce{OH+} and C, or \ce{OH} and \ce{C+}, can produce \ce{CO+}, which then forms \ce{HCO+}, followed by a dissociative electron recombination to \ce{CO}.

\subsubsection{Carbon chemistry}
High-energy radiation can influence the carbon chemistry in a few pronounced ways. First, the internally produced FUV radiation ionizes C to \ce{C+} and smooths out the transition regions between \ce{C+}/C/CO, producing broad ionization and dissociation fronts rather than the well-defined fronts in classical PDRs. This smoothening out can lead to more CO-dark gas \citep{Bisbas2015}, and enhances the amount of atomic carbon in dense gas. Second, the \ce{He+} can produce \ce{C+} in dense gas both through charge exchange with neutral carbon and through reactions with CO (Reaction eq:hepco). X-rays also directly photoionize carbon to produce single or doubly ionized carbon. At high temperatures, \ce{C+} hydrogenates through
\begin{equation}
\ce{C+ + H2 -> CH+ + H},
\end{equation}
although this reaction has an endothermicity of 0.4 eV \citep{Adams1984}. A faster reaction to form \ce{CH+} is through \ce{H3+},
\begin{equation}
\ce{H3+ + C -> CH+ + H2},
\end{equation}
which has an estimated Langevin reaction rate of $2\times10^{-9}$ cm$^3$ s$^{-1}$ \citep{Wakelam2024}. Following the production of \ce{CH+}, one can form other hydrocarbons
\begin{align}
\ce{CH+ + H2 &-> CH2+ + H} \\
&\rm{...} \nonumber\\
\ce{CH_n^+ + H2 &-> CH_{n+1}^+ + H}.
\end{align}
Other observationally important ions, such as \ce{HCO+}, are formed through proton exchange reactions,
\begin{equation}
\ce{H3+ + CO -> HCO+ + H2},
\end{equation}
or through reactions with gaseous water,
\begin{align}
\ce{C+ + H2O &-> H + HCO+} \\
\ce{&->H + HOC+},
\end{align}
with \ce{HOC+} slightly preferred over \ce{HCO+} at 10 K \citep{Wakelam2024}. There has also been laboratory work involving the particle-induced destruction and X-ray photodestruction of organics, such as acetic acid \citep{Boechat-Roberty2005, Pilling2006a, Pilling2006b, Pilling2011}, where the proton-impact destruction cross sections were measured. The focus on the destruction of complex organics by high-energy radiation has primarily been on the solid phase, where, in cold molecular gas, they are thought to form primarily.

\subsubsection{Nitrogen chemistry}
Nitrogen can be influenced through high-energy irradiation both through ion-neutral reactions and through the elevated gas temperatures these regions exhibit. \ce{N+} can be produced both through X-ray photoionization and secondary ionization and through destruction of \ce{N2} by \ce{He+}. Following the production of \ce{N+}, a series of reactions (the initial and final being slightly endothermic) leads to the formation of \ce{NH4+}, which then recombines to \ce{NH3}:
\begin{align}
\ce{N+ + H2 &-> NH+ + H} \\
\ce{NH+ + H2 &-> NH2+ + H} \\
\ce{NH2+ + H2 &-> NH3+ + H} \\
\ce{NH3+ + H2 &-> NH4+ + H} \\
\ce{NH4+ + e- &-> NH3 + H {\rm or} H2 + NH2 {\rm or} H + H + NH2}
\end{align}
Since both the start and end of this chain require reactions that are mildly endothermic, it is unclear how much of the ammonia is formed through this mechanism or via grain chemistry. The observationally important molecule \ce{N2H+} forms through proton exchange
\begin{equation}
\ce{H3+ + N2 -> N2H+ + H},
\end{equation}
which can also recombine to form \ce{N2}. 

When the gas is warmed by X-rays or cosmic rays, the ratio \ce{HNC/HCN} can be altered. In high ionization environments, these molecules are formed through \ce{HCNH+},
\begin{align}
\ce{HCNH+ + e- &-> HCN + H\\
           &-> HNC + H\\
           &-> CN + H2}
\end{align}
in rough equipartion. However, isomerization reactions can occur at higher temperatures
\begin{equation}
\ce{HNC + H -> HCN + H},
\end{equation}
which has an activation barrier exceeding 1000 K, although observations have indicated that either this reaction proceeds in a different manner, or there are other processes that regulate this ratio \citep{Hacar2020}, such as UV radiation \citep{Santa-Maria2023}.

\subsubsection{Deuteration}
Within the dense gas, deuteration can proceed following the formation \ce{HD}. \ce{HD} forms through the interaction of ionized deuterium with \ce{H2}. In cold gas, 
\begin{equation}
\ce{H3+ + HD -> H2 + H2D+}.
\end{equation}
The formation of \ce{H2D+} enables the inclusion of deuterium into more complex molecules in a similar manner as \ce{H3+}, i.e.,
\begin{equation}
\ce{H2D+ + CO -> DCO+ + H2}
\end{equation}
or 
\begin{equation}
\ce{H2D+ + N2 -> N2D+ + H2}.
\end{equation}
In this way, gas-phase deuteration can proceed in cold clouds, and has been produced as a chemical clock, with simpler species such as \ce{H2D+} being used as a tracer of the ionization rate \citep{Bovino2020}.

\subsubsection{Other chemistries of interest}
The chemistry of several other species are also intimately tied to high-energy radiation. There has been recent interest in the molecule argonium, \ce{ArH+}, due to its ubiquity in diffuse gas \citep{Schilke2014}. The chemistry of \ce{ArH+} is relatively simple, with both avenues tied directly to high-energy radiation. First, if argon is ionized directly, it can either recombine through electron recombination or grain-assisted recombine, or it can react with \ce{H2},
\begin{equation}
   \ce{Ar+ + H2 -> ArH+ + H}. 
\end{equation} 
\ce{ArH+} can also form through \ce{H3+},
\begin{equation}
\ce{Ar + H3+ -> ArH+ + H2}.
\end{equation}
\ce{ArH+} undergoes electron dissociative recombination, but also acts as a proton donor, i.e., \ce{ArH+ + CO -> Ar + HCO+}. Due to its chemistry being relatively simple and directly tied to high-energy radiation, it can be a sensitive probe of the ionization rate in the diffuse interstellar medium \citep{Schilke2014, Priestley2017}.

Recently, there has been a sizable interest in the chemistry of phosphorus-bearing molecules in the interstellar medium, in particular PO and PN. These molecules have been detected now across the range of molecular cloud environments \citep{Turner1987, Ziurys1987, Fontani2016, Lefloch2016, Rivilla2016, Rivilla2018, Fontani2019, Scibelli2025}, including a confirmed extragalactic detection \citep{Haasler2022}. The start of the chemistry is thought to originate from phosphorus being kicked off grains, primarily by shocks. For regions with strong shocks or hot gas, neutral-neutral reactions have been found to be plausible, \ce{PN + O} \citep{Jimenez-Serra2018}, \ce{P + OH} \citep{GarciadelaConcepcion2021} or \ce{P + O2} \citep{GarciadelaConcepcion2024}. However, the molecule \ce{PO+} has been discovered both in the galactic center \citep{Rivilla2022} and in starless cores \citep{Scibelli2025}. Here, the current proposal is that \ce{P} is directly ionized through cosmic-ray processes followed by reaction with OH, e.g., \ce{P+ + OH -> PO+ + H}. Furthermore, \ce{PO} and \ce{PN} can be directly ionized, although there is currently a lack of reaction rates for these species. The ratio $N(\ce{PO+})/N(\ce{PO})$ has been proposed as a possible tracer for energetic radiation \citep{Rivilla2022}.

The molecule \ce{SH+} is also formed in part through ionization processes \citep{Wolfire2022}. Internally generated FUV radiation dissociates neutral \ce{SH}, producing neutral atomic sulfur. Atomic sulfur is readily ionized by X-ray radiation or cosmic rays, which then reacts with neutrals to form \ce{SO+} or \ce{SH+}. Reactions with neutral sulfur with, e.g. \ce{H3+}, leads to \ce{SH+}. 

It is worth emphasizing that for X-ray ionization processes, and to an extent, cosmic-ray irradiation, much of the direct ionization impact is through secondary electrons. Historically, there has been a minimal inclusion of the electron-impact ionization from secondary electrons, although reaction networks have included the induced CR or X-ray FUV radiation. The new reaction rates and cross sections calculated by \citet{Gaches2024} will expand the availability of ionization processes due to the availability of these cross-section data for hundreds of molecules. As such, the gas-phase chemistry in the molecular interstellar medium is still not a solved problem, with crucial fundamental data still being generated or not yet available. 

\subsection{Gas heating via high-energy radiation}
There are some similarities, but also key differences, in how heating and cooling are treated between cosmic ray and X-ray irradiation. The heating induced by X-rays and cosmic rays has been extensively investigated \citep{Dalgarno1999, Meijerink2005, Glassgold2012}. For X-ray processes, a parameter of fundamental importance is the energy deposition rate per particle, $H_X$,
\begin{equation}
H_X = \int_{E_{\rm min}}^{E_{\rm max}} \sigma_{pa}(E)F_X(E)dE.
\end{equation}
Since molecular processes and cooling scale as roughly $n^2$, and the total heating rate is $nH_X$, the influence of X-rays is often described through $H_X/n$.

The volumetric heating rates (erg cm$^{-3}$ s$^{-1}$) are characterized similarly between cosmic rays and X-rays. For cosmic rays,
\begin{equation}
\Gamma_{\rm CR} = Q n \zeta_H,
\end{equation}
where $Q$ is the average energy deposited in the gas per ionization. The physics that go into $Q$ are complicated, but recent calculations by \citet{Glassgold2012} provide $Q = 4.3$ eV in atomic gas and $Q \approx 10 - 17$ eV in dense gas, depending on the density.

For X-ray heating, there is a similar expression,
\begin{equation}
\Gamma_X = \eta n H_X,
\end{equation}
where $\eta$ is the heating efficiency of the X-ray radiation. The efficiencies have been computed by detailing all possible loss processes (see below) by both \citet{Dalgarno1999} and \citet{Glassgold2012} and depend both on the typical photon energy, the relative amount of H, \ce{H2}, and He, and the ionization fraction.

Gas heating is brought on by a combination of processes: elastic collisions, rovibrational \ce{H2} excitations, \ce{H2} dissociation, and chemical heating. The canonical work of \citet{Dalgarno1999} presented the calculations of the collisional processes, while \citet{Glassgold2012} expanded this to include chemical heating, which they found significantly increased the deposited heat per ionization. There have also been recent calculations of energy deposition of heating of cosmic-ray primary particles and induced secondaries using spherically-symmetric clouds and molecular clouds in simulations using the Geant4 physics package, which allowed the investigation of the deposition due to a variety primary ions and induced secondaries \citep{Pazianotto2021, Pilling2021, Pilling2022b, Pazianotto2023}. These works have not compared their calculated energy depositions to the per-ionization-event heating rates of previous works that are typically used in thermodynamical and chemical models of molecular clouds, mentioned above.

The overall heating rate is sensitive to the gas composition, temperature, and ionization fraction, making a proper calculation of the thermodynamics and chemistry of these regions complicated, especially if atomic and molecular level populations are not assumed to be in thermal equilibrium. Heating is more efficient in molecular gas due to the increased number of energy loss channels provided by \ce{H2} (electronic, vibrational, and rotational) compared to atomic hydrogen (electronic). In atomic gas, the heating efficiency is around 10\%, and in molecular gas it can exceed 50\% \citep{Glassgold2012}. In dense molecular gas, the gas temperature is directly related to the ionization rate, with gas temperatures exceeding 30 K possible for ionization rates $\zeta > 10^{-15}$ s$^{-1}$ \citep{Bisbas2015}.

\begin{figure*}
    \centering
    \subfloat[$n_H = 100$ cm$^{-3}$]{\includegraphics[width=0.33\textwidth]{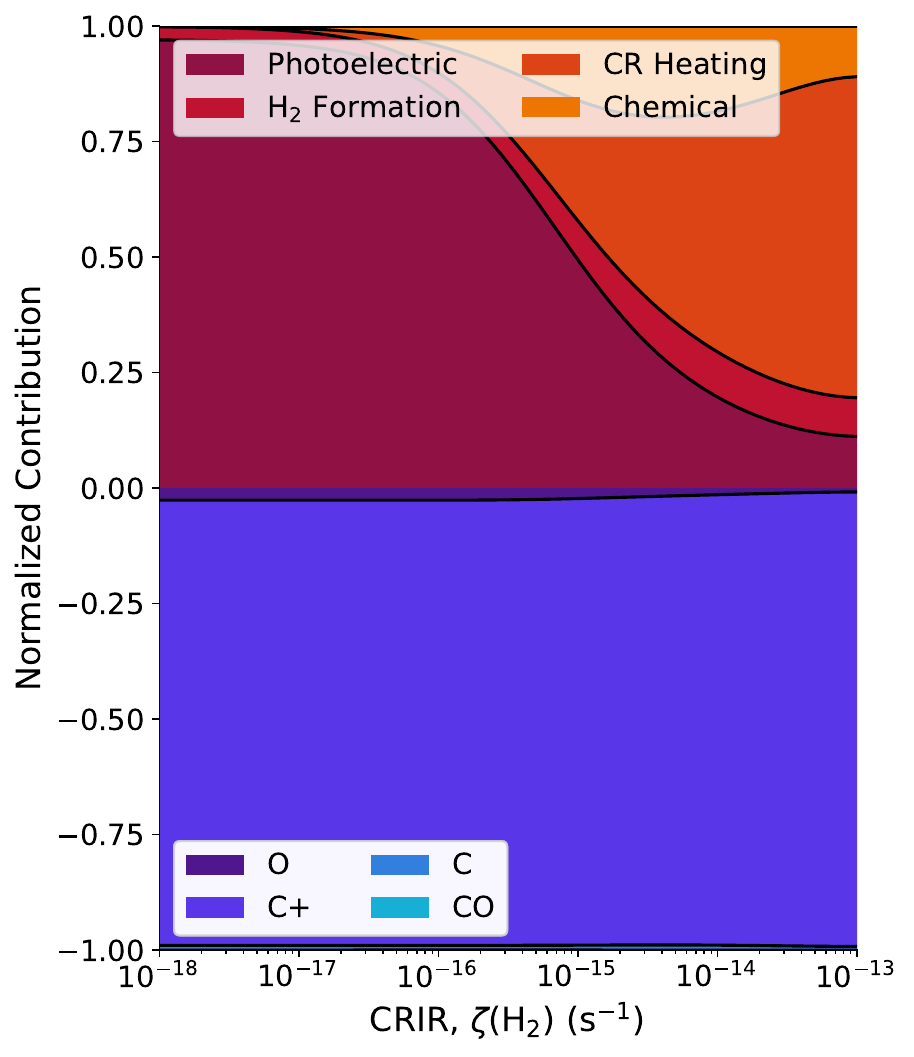}}
    \subfloat[$n_H = 10^3$ cm$^{-3}$]{\includegraphics[width=0.33\textwidth]{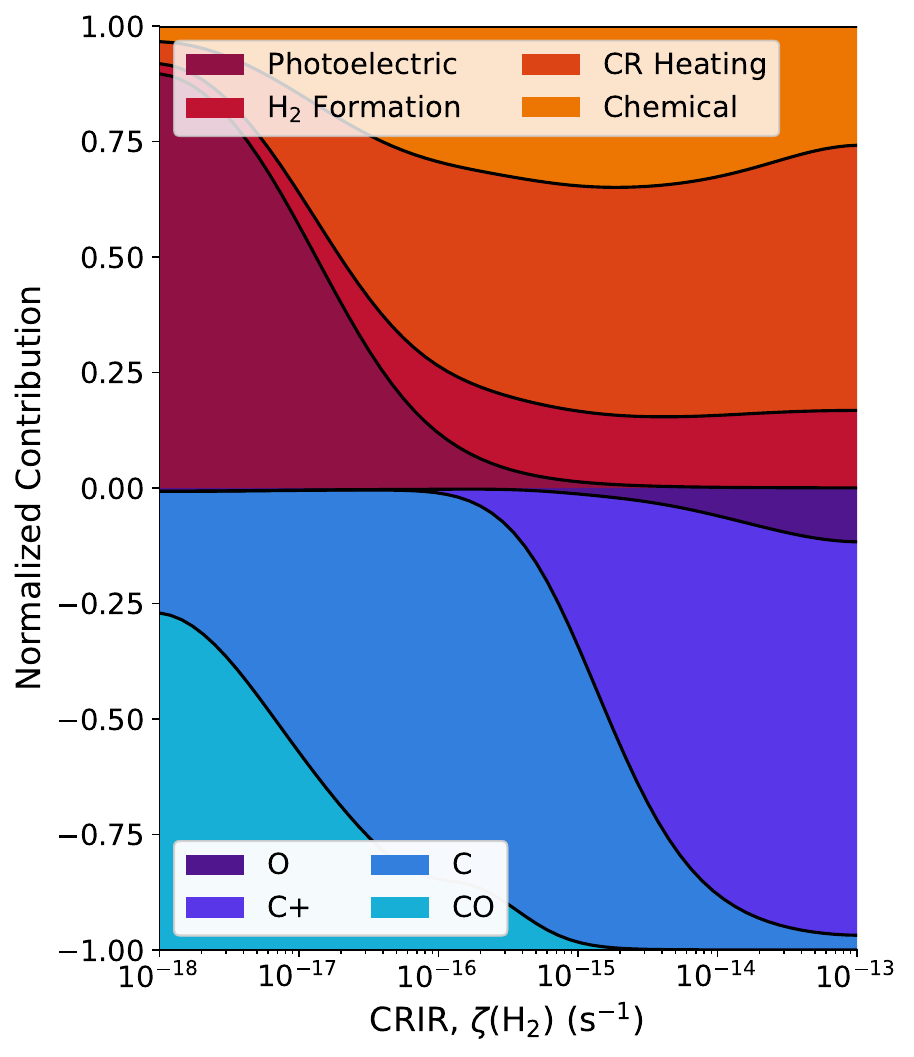}}
    \subfloat[$n_H = 10^4$ cm$^{-3}$]{\includegraphics[width=0.33\textwidth]{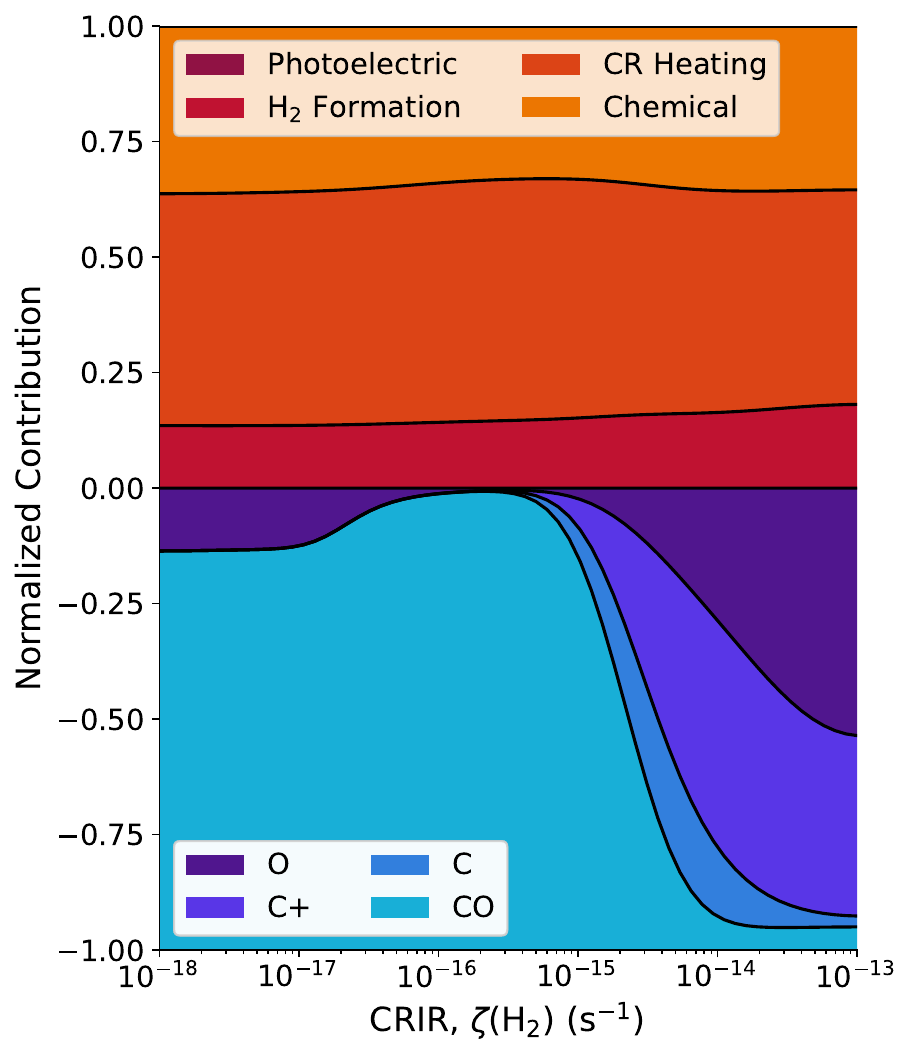}}
    \caption{\label{fig:heatcoolcontrib}The normalized contributions of different heating (top) and cooling mechanisms (bottom) for three different densities, $n_H = 10, 10^3, 10^4$ cm$^{-3}$ (a, b, c, respectively) as a function of \ce{H2} cosmic-ray ionization rate.}
\end{figure*}

While the overall heating is quite similar in individual terms, a major difference between cosmic-ray-driven processes and X-ray irradiation involves the initial production, and thus the spectrum, of fast electrons. While the secondary spectrum has been recently calculated in detail for cosmic rays \citep{Ivlev2021}, a similar analysis has not been done yet for X-rays. Secondary electron production from X-rays is a combination of primary photoelectrons from H, \ce{H2} and He for soft X-rays, and for hard X-rays, the photoabsorption cross section is dominated by carbon, oxygen, and nitrogen (and other metals\footnote{It is worth reminding readers that in astronomical literature, all elements heavier than helium are called metals.}). X-ray photoelectric emission is prone to producing an Auger cascade, enhancing the amount of fast electrons ejected into the gas per primary ionization event. While proton- and electron-impact ionization can produce multiple ionizations, and Auger ionizations for inner electron orbitals of molecules with heavy atoms, the inclusion of these multiples is only a minor contribution to the overall ionization cross section, indicating single ionization is dominant for most species of astrochemical interest \citep{Nishimura1999}. There is also a difference between X-rays and cosmic rays in the depths that are able to be reached, with cosmic rays able to penetrate significantly further into molecular gas.

Compared with the classical PDR models, XDRs exhibit noticeably different cooling. Since X-rays couple tightly to the gas, much of the heating goes directly into the gas rather than the dust. Therefore, there is a significant component of cooling through atomic and molecular lines. In the warmest regions, there is a substantial amount of atomic line cooling from infrared atomic lines, [O I] 63$\mu$m, [Si II] 35 $\mu$m, and [CII] 158 $\mu$m \citep{Maloney1996}. Furthermore, secondary ionizations of metals can provide ionic cooling lines, in particular [Si II]. Further into the clouds, the cooling becomes dominated by neutral carbon lines at [C I] 370$\mu$m and 609$\mu$m. In the densest molecular regions, the gas can still be substantially warm, leading to a high flux of high-J CO lines in the THz. Finally, X-ray and cosmic-ray heating can lead to \ce{H2} warm enough to cool via rovibrational lines. Given the wavelength regimes where line cooling is important, the mid- to far-infrared lines can act as a sensitive probe to the underlying radiation fields in the molecular gas. However, at the time of writing, there is currently a stark lack of observatories sensitive to the far infrared (FIR) and THz lines, although higher frequency bands of the Atacama Large Millimeter Array (ALMA) or the proposed Atacama Large Aperture Submillimeter Telescope (AtLAST) telescope \citep{Booth2024}, along with observations from the James Webb Space Telescope, will be crucial for constraints on the high-energy radiation.

Figure \ref{fig:heatcoolcontrib} shows the normalized contributions to the total heating and cooling functions for various components for cosmic-ray irradiation. The sub-figures show the model results (see below for details) for densities $n_H = 100, 10^3$ and $10^4$ cm$^{-3}$, corresponding to attenuating extinctions, $A_V \approx 0.65, 2,$ and $7$ mag, as a function of cosmic ray ionization rate, with a background radiation field corresponding to the solar neighborhood interstellar radiation field. The figure demonstrates what is described above. In low densities, the heating is dominated by the photoelectric heating until $\zeta(\ce{H2}) \approx 5\times 10^{-16}$ s$^{-1}$, while the cooling remains primarily dominated by \ce{C+}. At intermediate densities, $10^3$ cm$^{-3}$, the photoelectric heating is only dominant until $\zeta > 10^{-17}$ s$^{-1}$, while for higher ionization rates it is split between direct CR heating and chemical heating (itself primarily a result of cosmic-ray initiated ion-neutral reactions). For the cooling, at low CRIR, it is dominated by CO, with C taking over between $10^{-17} \leq \zeta(\ce{H2}) \leq 10^{-15}$, followed then by \ce{C+}. At high densities, only cosmic-ray and chemical heating are significant, while CO becomes a strong coolant, followed by \ce{C+} and O. It is worth noting that the temperature here is highest due to the elevated temperatures.

\begin{figure*}
    \centering
    \includegraphics[width=\textwidth]{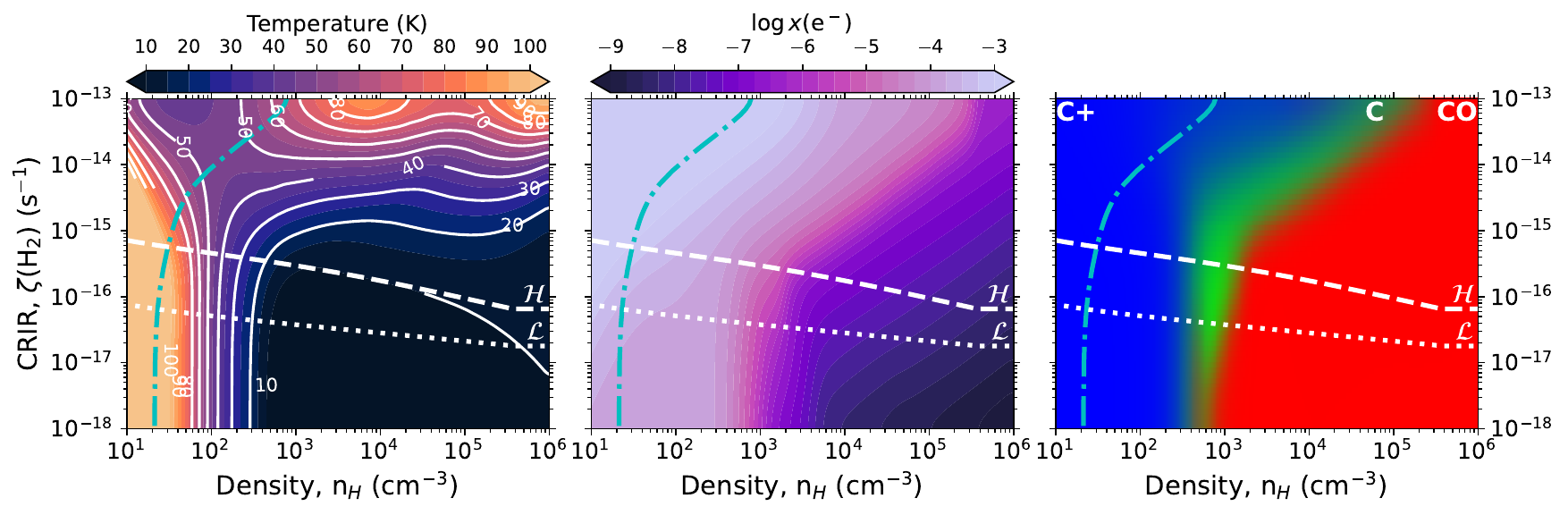}
    \caption{\label{fig:crchem}Left: Gas temperature, $T_{\rm gas}$ as a function of density and total \ce{H2} CRIR, $\zeta(\ce{H2})$. Center: Same as left, but showing the electron fraction, $x(\ce{e-})$. Right: Same as left, but showing a three-color image for the abundance of \ce{C+} (blue), C (green), and CO (red). The cyan dashed-dotted line denotes where $x(H) = 2x(H_2)$. The annotated white dashed and dotted lines show the cosmic ray attenuation models of \citet{Padovani2018}, denoted $\mathcal{H}$ and $\mathcal{L}$.}
\end{figure*}

\subsubsection{Demonstration through PDR/CRDR grid}
The impact of increasing the cosmic-ray ionization rate is demonstrated in Figure \ref{fig:crchem}, which utilizes a grid of one-dimensional density distributions following the fitted relationship between the effective attenuation extinction, $A_{V, {\rm eff}}$ and number density, $n_H$, from \citet{Bisbas2023} (see also \citep{Gaches2025} for analytic models),
\begin{equation}
    A_{V, {\rm eff}} = 0.05 \exp{\left [ 1.6 \left ( \frac{n_H}{{\rm cm^{-3}}} \right )^{0.12} \right ]} ~{\rm mag}.
\end{equation}
The relationship produces a density distribution in which low-density gas has low $A_V$ and high-density gas has high $A_V$, and has provided a good approximation to mean trends found in three-dimensional simulations \citep{Bisbas2023}. The models are calculated using the public photo-dissociation region code \textsc{3d-pdr} \citep{Bisbas2012}\footnote{\url{https://uclchem.github.io/3dpdr}} with an initial atomic composition matching the solar neighborhood and a solar neighborhood external FUV radiation field \citep{Draine1978}. The models only include gas-phase chemistry (except for \ce{H2} formation), but fully solve the thermodynamics and non-thermal level populations for \ce{C+}, C, O, and CO. The figure also shows two cosmic ray attenuation models from \citet{Padovani2018}, $\mathcal{H}$ and $\mathcal{L}$, where the former was derived to reproduce diffuse gas measurements of $\zeta(\ce{H2})$ and the latter is consistent with the cosmic ray spectrum measured by Voyager, using the relationship between the density and attenuating column density. 

Figure \ref{fig:crchem} shows that above a CRIR, $\zeta(\ce{H2}) \ge 10^{-16}$, the gas temperature is primarily determined by the CRIR. Further, the electron fraction increases with CRIR, although there is still considerable density dependence. Regions with extremely high ionization rates, $\zeta > 10^{-14}$, can maintain ionization fractions of $x(\ce{e-}) \approx 10^{-4}$ even in denser gas. Finally, the sub-panel c shows a representation of the carbon cycle, \ce{C+}/C/CO in blue/green/red, and highlights that for increased CRIRs, $\zeta(\ce{H2})$, CRs can start to destroy significant amounts of CO leading to atomic (neutral and ionized) carbon in dense gas with $n_H > 10^3$ cm$^{-3}$. At $\zeta(\ce{H2}) > 10^{-15}$ s$^{-1}$, Figure \ref{fig:crchem} shows that the H/\ce{H2} transition moves deeper into the cloud to higher densities as the cosmic rays finally begin to impact the abundance of molecular hydrogen appreciably. Since these are one-dimensional with an interstellar radiation field, much of the gas temperature and ionization for $n_H < {\rm few} \times 100$ cm$^{-3}$ is caused by the ISRF's FUV flux.

\subsubsection{Demonstration through XDR grid}
We present here a demonstration of the impact of X-ray fluxes on the thermochemistry in dense molecular gas. We utilize the {\sc Cloudy}\footnote{\url{https://gitlab.nublado.org/cloudy/cloudy}} \citep{Gunasekera2025} spectral synthesis code to run a grid of slab models in a range of constant densities, $100 < n_H < 10^5$ cm$^{-3}$ and total integrated X-ray fluxes, $10^{-4} < F_X < 100$ erg cm$^{-2}$ s$^{-1}$. We include a background interstellar radiation field, the cosmic microwave background, and a cosmic-ray ionization rate, $\zeta_H = 10^{-20}$ s$^{-1}$. We utilized the ``XDR.sed'' SED table, which takes the form
\begin{equation}
    F(\nu) = F_0 \times \left ( \frac{E}{100 ~{\rm keV}}\right )^{-0.7} ~{[\rm erg \, cm^{-2}\, s^{-1}\, Hz^{-1}]}.
\end{equation}
We run the slabs until an extinction, $A_V = 5$ mag, and all abundances and gas temperatures are plotted at this point. 

\begin{figure*}
    \centering
    \includegraphics[width=\textwidth]{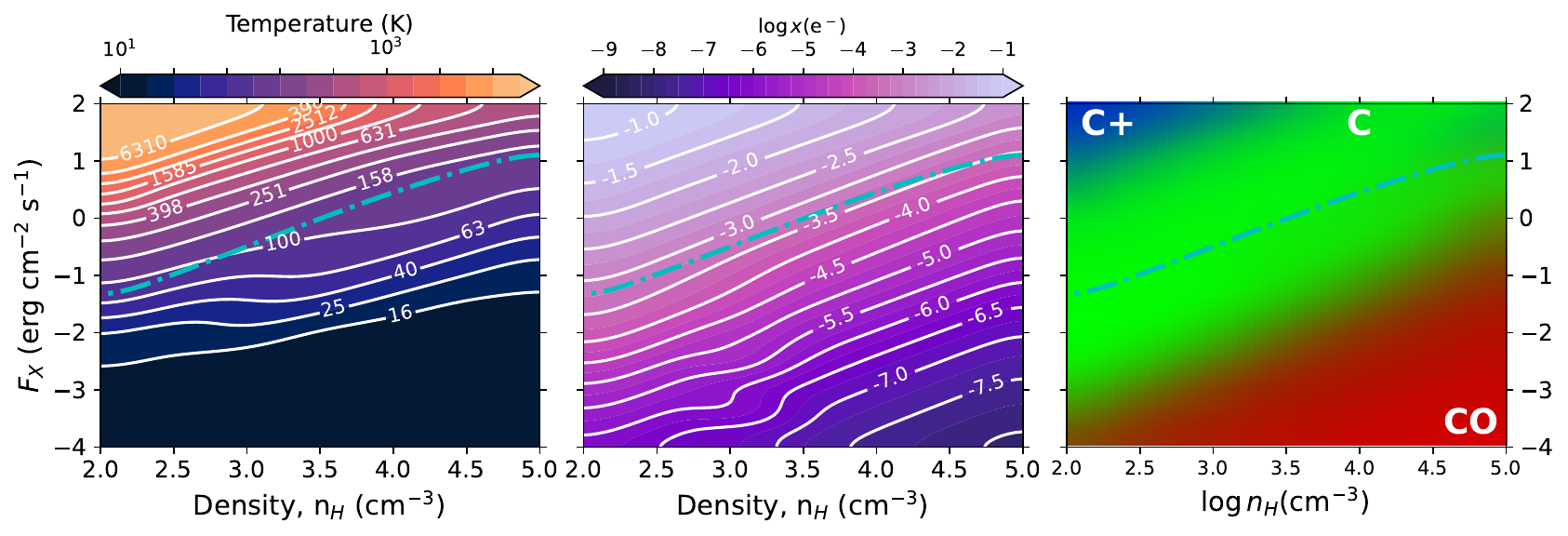}
    \caption{\label{fig:XDR}Left: Gas temperature versus number density, $n_H$ (cm$^{-3}$), and integrated X-ray flux, $F_X$ (erg cm$^{-2}$ s$^{-1}$). Center: Same as left, but showing the electron fraction, $x(\ce{e-})$. Right: Same as left, but showing a three-color image for the abundance of \ce{C+} (blue), C (green), and CO (red). The cyan contour shows where $x(\ce{H}) = 2x(\ce{H2})$.}
\end{figure*}

Figure \ref{fig:XDR} shows that the X-rays efficiently heat the gas at $n_H > 100$ cm$^{-3}$ and significantly the H-\ce{H2} transition point to higher densities. The gas for modest integrated X-ray fluxes heats to tens to a hundred Kelvins. Since the model has a weak FUV field and a vanishing CRIR, the electron fraction is determined solely by the X-ray flux and density. The results agree with the {\sc 3d-pdr} models: in low X-ray fluxes in dense gas, the electron fraction approaches $x(e) \approx 10^{-8}$. For intense fields, the gas starts to become fully ionized. The electron function scales with the X-ray flux and inversely with the density, although weaker than the dependency on the X-ray flux. There is a significant amount of the parameter space where carbon is primarily in the neutral atomic phase. For strong X-ray fluxes, ionized carbon can exist in gas with $n_H > 100$ cm$^{-3}$, while CO only exists in denser regions with reduced X-ray fluxes, $F_X < 10^{-2}$ erg s$^{-1}$ cm$^{-2}$.

\section{Energetic ice processing}\label{sec:icephase}
High-energy radiation, in the form of X-rays and cosmic rays (and other energetic particles), highly processes the ices on grains, including both their chemical composition and their morphological structure. A plethora of laboratory studies have demonstrated the feasibility of high-energy radiation to synthesize complex organic chemistry from simple ice compositions. Laboratory investigations have found that both X-rays and high-energy particles, including heavy ions, act in similar underlying mechanisms in ice processing, via the production of secondary electrons and the resulting cascade \citep{Mason2014, Oberg2016, Vasconcelos2017, Arumainayagam2019}. The ice chemistry induced by high-energy radiation is, by definition, radiation chemistry (radiolysis) and progresses through ionization and excitation processes.

\subsection{Modelling framework}
Before detailing the chemistry, we first describe some of the fundamental terminology. \citet{Shingledecker2018a} presented a model for radiation chemistry induced by cosmic rays (and here unified with hard X-rays) in ices using the commonly-used G-values for yields, associated with the stopping cross sections (also called loss functions), $S$ and the $W$ values, described above. Here, we follow this notation for both X-ray and cosmic-ray induced chemistry in ices. Following \citet{Shingledecker2018a}, we denote four primary types of reactions, where suprathermal species in excited states are denoted with an asterisk, and we utilize the symbol $\leadsto$ to denote action by either the primary ionizing radiation or secondary electrons,
\begin{align}
\ce{X_s &\leadsto X_s+ + e-} \label{eq:sion1} \\
\ce{X_s &\leadsto X_s+ + e- -> X_s^* -> Y_s^* + Z_s^*} \label{eq:sion2}\\
\ce{X_s &\leadsto X_s^* -> Y_s + Z_s} \label{eq:sexc1}\\
\ce{X_s &\leadsto X_s^*} \label{eq:sexc2}
\end{align}
and we extend this to include any desorption schemes
\begin{align}
\ce{X_s &\leadsto X_g} \label{eq:des1} \\
\ce{X_s &\leadsto Y_g + Z_g} \label{eq:des2},
\end{align}
where we have denoted examples species X, Y, Z in the solid phase, denoted by s, or gas phase, denoted by g. In the bulk, Reaction \ref{eq:sion1} can be a minor contribution since molecular ions quickly recombine or interact with nearby species. Due to this, we only show the G-value for ionization reaction \ref{eq:sion2}, which ends by dissociative electron recombination. Reactions \ref{eq:sexc1} and \ref{eq:sexc2} correspond to suprathremal excitation to a dissociative or stable state, respectively. The G-values, which describe the total number of species created or destroyed per 100 eV deposited, are as follows: \\
for Reaction \ref{eq:sion2}
\begin{equation}
G_i = \frac{100 {\,\, \rm {eV}}}{W},
\end{equation}
for Reaction \ref{eq:sexc1},
\begin{equation}
G_{de} = P_{\rm diss} \left ( \frac{100 {\,\, \rm {eV}}}{W}\right )\xi,
\end{equation}
and for Reaction \ref{eq:sexc2},
\begin{equation}
G_{e} = (1 - P_{\rm diss}) \left ( \frac{100 {\,\, \rm {eV}}}{W} \right ) \xi,
\end{equation}
where $P_{\rm diss}$ is the dissociation probability from state $X_s^*$, $\xi$ is the ratio of excitation to ionizing collisions,
\begin{equation}
\xi = \frac{W - E_{\rm ion} - W_s}{W_{\rm exc}},
\end{equation}
$E_{\rm ion}$ is the ionization energy of the species, $W_s$ is average sub-excitation energy and $W_{\rm exc}$ is the average excitation energy. At high energy, the G-values will be roughly consistent between the different sources of radiation, and so for each process, $i$, and each source of radiation, $k$, the reaction rate can be described by
\begin{equation}
k_{i,k} = G_i \left ( \frac{S_{e,k}}{100~{\rm eV}} \right ) \phi_k,
\end{equation}
where $S_{e,k}$ is the electronic loss function in the material for radiation source, $k$, and $\phi_k$ is the number flux of high-energy radiation of type $k$. The loss functions can be computed using software such as the SRIM\footnote{\url{http://www.srim.org/}} \citep{Ziegler2010} or PSTAR\footnote{\url{https://physics.nist.gov/PhysRefData/Star/Text/PSTAR.html}} for particle radiation or the CXRO X-ray Interactions with Matter tool\footnote{\url{https://henke.lbl.gov/optical_constants/}} \citep{Henke1993}. The total reaction rate is thus,
\begin{equation}
k_i = G_i \sum_{k = \gamma, {\rm particles}}\left ( \frac{S_{e,k}}{100~{\rm eV}} \right ) \phi_k.
\end{equation}
We note that the above unification is true for high energy X-rays when the yields become relatively constant (above, e.g., 200 eV \citep{Dalgarno1999}). Below this limit, the above reaction rates must be rewritten in terms of the spectrum and the photo-absorption cross sections \citep{Mullikin2021}.

When molecules are released from the ice, it is a process called desorption. In warm gas, this is dominated by thermal desorption processes. Energetic radiation also induces desorption processes, both from the induced stochastic heating and from direct non-thermal processes. Following \citet{Hasegawa1993}, the high-energy radiation-induced thermal desorption rate, for the case of cosmic rays, is
\begin{equation}
k_{crd, i} = 3.16 \times 10^{-9} k_{td,X}(70~{\rm K}),
\end{equation}
where the prefactor was computed assuming Fe nuclei dominate the collisional heating, $k_{td,X}$ is the thermal desorption rate,
\begin{equation}
k_{td,X} = \nu_{0,X}e^{-E_{D,X}/T_d},
\end{equation}
$\nu_{0,X}$ is the trial frequency, which is associated with the diffusion of the species, $E_{D,X}$ is the binding energy of the species onto the ice and $T_d$ is the resulting dust temperature. In many applications, it is broadly assumed that Fe nuclei dominate these collisional processes since the electronic stopping power scales with $Z^2$ \citep{Dartois2023}. There has been a significant amount of work, in particular by Kalv\={a}ns and collaborators \citep{Kalvans2016, Kalvans2018, Kalvans2022} in accurately computing the dust temperatures following cosmic ray impact. A similar formalism has not, to our knowledge, been carried out for X-ray irradiation, although \citet{Yan1997} gives the following expression for dust temperature as a function of $H_X$,
\begin{equation}
T_d = 1.5 \times 10^4 \left (\frac{H_X}{x_d} \right )^{0.2} ~{K},
\end{equation}
where $x_d$ is the grain abundance.

Furthermore, as discussed above, the ionizing radiation produces an embedded source of FUV radiation which can process the ice, with a reaction rate \citep{Ruaud2016},
\begin{equation}
k_{\rm des, UVCR} = \phi_{\rm UVCR} S_{\rm UVCR} Y_{\rm pd} \left ( \frac{\pi r_{\rm dust}^2}{N_{\rm site}} \right ),
\end{equation}
where $\phi_{\rm UVCR}$ is the flux of secondary UV photons, $S_{\rm UVCR}$ is a scaling factor that can employed, $r_{\rm dust}$ is the dust size, $Y_{pd} = 10^{-4}$ molecules/photon is the assumed desorption yield for this band, and $N_{\rm size}$ is the number of surface binding sites on the grain. 

For X-ray radiation, the total desorption rate of species, $i$, can be described \citep{Walsh2010, Walsh2012} by
\begin{equation}
k_{i}^{X} = F_X Y_X P_{\rm abs} \sigma_d x_d ~{\rm s^{-1}},
\end{equation}
where $Y_X$ is the total yield of desorbed molecules, $P_{\rm abs}$ is the probability of absorption onto the grain and $\sigma_d$ is the total dust geometric cross section (and thus $P_{\rm abs}\sigma_d$ is the effective absorption cross section). The above relation holds if the yield, $Y_X$, is calculated for a spectrum associated with the X-ray source. Otherwise, a spectrum-averaged rate can be computed with energy-dependent yield functions,
\begin{equation}
k_i^X = P_{\rm abs}\sigma_d x_d \int_{E_{\rm min}}^{E_{\rm max}} F_X(E)Y_X(E) dE ~{\rm s^{-1}},
\end{equation}
where it is assumed that $P_{\rm abs}$ does not appreciably change over energy. The yields have been calculated by an increasing number of experiments for a range of molecules, ice mixtures, photon fluences, and irradiation energies \citep{Andrade2010, Jimenez-Escobar2012, Jimenez-Escobar2016, Dupuy2018, Jimenez-Escobar2018, Ciaravella2019, Basalgete2021a, Basalgete2021b, Dupuy2021, Carvalho2022, Basalgete2023, Carvalho2024, Torres-Diaz2024}.

\begin{figure}
    \centering
    \includegraphics[width=\columnwidth]{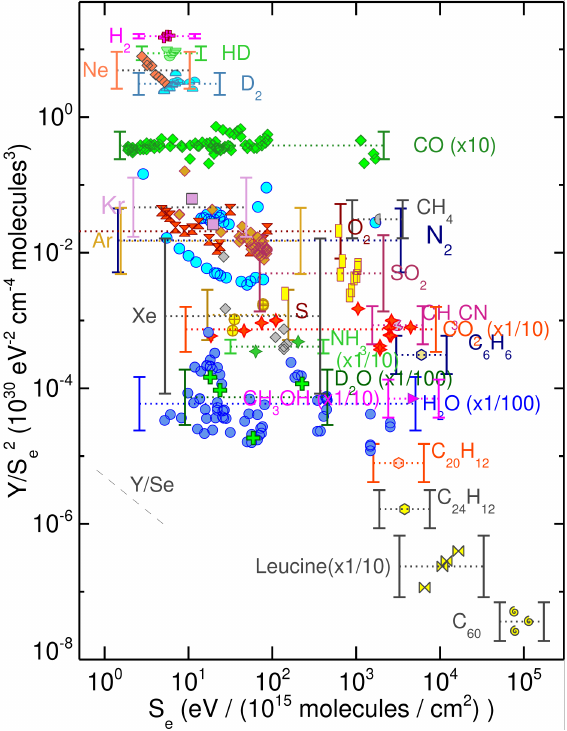}
    \caption{\label{fig:dartois} Compilation of sputtering yields, $Y$, normalized by the quadratic of the electronic stopping power $S_e^2$, versus $S_e$. The bars show the computed error of the sputtering prefactor. Reused with permission from \citet{Dartois2023}.}
\end{figure}

Finally, heavy ion irradiation can directly sputter material off of the grain surface. We emphasize the role that heavy ions play in this, due to the scaling of the sputtering rate with $Z^4$ \citep{Dartois2023}. Following \citet{Dartois2023}, the sputtering yield per ion is written as
\begin{equation}
Y^{\rm tot} \approx Y_n^0 S_n + Y_e^0 S_e^2,
\end{equation}
where $Y_n^0$ and $Y_e^0$ are the yield prefactors that can be determined for materials from laboratory experiments. In general, for systems of astrochemical interest, $Y^{\rm tot} \approx Y_e^0 S_e^2$. The sputtering yields in the electronic regime were recently brought together and compiled for many molecules, including complex organics, in \citet{Dartois2023}. Figure \ref{fig:dartois} shows compiled sputter yields from \citet{Dartois2023} normalized by $S_e^2$ for different molecules versus $S_e$. These yields are consistent with $Y \propto S_e^2$. Some molecules show discrepancies, such as Leucine (\ce{C6H13NO2}) and \ce{N2}, but are still consistent with the above scaling within the computed error of the prefactor. \citet{Dartois2023} showed that to first order, $Y^{\rm tot} \propto \Delta H^{-2.2 \pm 0.6}_{\rm sub}$ when excluding the most massive molecules in their dataset, where $\Delta H$ is the sublimation enthalpy. The total sputtering rate, for galactic cosmic rays, is
\begin{equation}
k_{\rm CR,sput} = 4\pi\sum_Z \int_{E_{\rm min}}^{E_{\rm max}} Y^{\rm tot}(E, Z) j_Z(E) dE,
\end{equation}
where the energy dependence of $Y^{\rm tot}$ comes from the stopping cross section. For ionization rates of $O(10^{-17}$ s$^{-1})$, small molecules have comparable sputtering rates to the secondary FUV desorption rate, but for larger molecules, the sputtering rate can exceed the secondary FUV desorption rate by factors of a few to a hundred \citep{Dartois2023}.

\subsection{Radiation chemistry in the ice}
The chemistry in the ice, with the exception of sputtering, is driven through stochastic spot heating, leading to a rapid thermal chemical processing, and the sizable production of induced secondary electrons. The secondary electrons drive chemistry through Reactions \ref{eq:sion2} - \ref{eq:sexc2}, with both ionizing and low-energy electrons playing important roles. In recent years, it has become broadly accepted that the chemistry is determined by the flux of low-energy electrons with energies $\lesssim$ 20 eV \citep{Boyer2016, Arumainayagam2019}. In fact, for the radiolysis, what is most important to the chemistry is the total amount of ionizing energy that is input into the system, regardless of whether it is caused by energetic particles or high-energy photons \citep{Oberg2016, Arumainayagam2019}. What separates the VUV/EUV photochemistry from X-ray and cosmic-ray driven chemistry is that the latter forms of radiation induce substantial secondary electron cascades and penetrate much further into the ice and grain \citep{Andrade2008}. Collisions with energetic particles can even impact the underlying mineral grain substrate, impacting the heating \citep{Ivlev2015a, Kalvans2018, Kalvans2022} and charge \citep{Ivlev2015b}.

There have been a plethora of laboratory experiments investigating the radiolysis of astrophysical ices, with experiments of X-ray radiation, electrons, and heavy ions now providing a significant amount of valuable information on the induced chemistry. There are now a growing number of sophisticated experimental setups for high-energy radiolysis, such as INFRA-ICE \citep{Santoro2020}, Ice Chamber for Astrophysics-Astrochemistry (ICA) \citep{Herczku2021}, the Versatile Ice Zigzag Sublimation Setup for Laboratory Astrochemistry (VIZSLA) \citep{Bazso2021}, AQUILA \citep{Racz2024}, the Brazilian Synchrotron Light Laboratory (LNLS), and the Grand Acc\'{e}l\'{e}rateur National d'Ions Lourds (GANIL), with the latter facility now operating the SPIRAL2 LINAC, capable of accelerating light nuclei. These experiments use either infrared spectra or quadrupole mass spectrometry (QMS) data with temperature-programmed desorption (TPD) to detect the products of irradiation. In addition, both pure ices and mixed ices have been investigated, with the latter including mixed ices with nitrogen (typically through \ce{NH3}) that are vital to constrain pathways towards prebiotic chemistry. Elucidating the exact radiolysis pathways can be tricky, although in some cases effective reaction rates can be fitted to laboratory data \citep{Carvalho2022, Pilling2022, Carvalho2024, daSilveira2024}. Despite these difficulties, various aspects of the underlying radiolysis chemistry in the bulk have been well constrained for systems common in astrophysical ices, as detailed below. 

The induced chemistry in the bulk is dominated by the interactions of low-energy electrons, produced by the primary X-ray photon and cosmic ray particles. During the primary interaction, the dominant interaction results in ionization, with cosmic ray ions primarily leading to single ionizations and high-energy X-rays leading to multiple ionizations due to the Auger effect. These secondary electrons can further ionize, resulting in a cascade ending with a large population of low-energy electrons. \citet{Shingledecker2020} simulated the collision of cosmic-ray protons with energies between 100 keV and 100 MeV through a 1 $\mu$m thick block of water ice with the Geant4-DNA physics package \citep{Incerti2018}. The simulations provided a first-principles calculation of the ionization track, finding that the secondary electron distribution is well-described by a cylinder around the primary interaction track. The radius of the cylinder was weakly energy dependent until $\approx$ 5 MeV after which it is constant, with an inverse dependence on the ice density. Future simulations of the interactions between high-energy radiation and ice, such as these, will be greatly helpful in elucidating the microphysics of their interactions.

The ionizations drive an initial ion-neutral chemistry with cations, and both the ionizations and initial thermal deposition of the primary radiation lead to sputtering off the surface of fragments. The induced low energy ($<$20 eV) further stimulates a complex chemistry in the bulk of the gas through excitations and dissociative electron attachment (DEA). For instance, following irradiation\citep{Arumainayagam2019}, water ice can undergo a wide range of processing, from ionization
\begin{align}
    \ce{H2O &\leadsto H2O+ + e-} \\
    \ce{H2O+ + H2O &-> H3O+ + OH}
\end{align}
or at lower energies, excitation
\begin{align}
    \ce{H2O \leadsto H2O^* &-> H2O + h\nu} \\
    \ce{& -> H2 + O} \\
    \ce{& -> H + H + O} \\
    \ce{& -> OH + H}
\end{align}
or DEA,
\begin{align}
    \ce{H2O \leadsto H2O- &-> H- + OH} \\
    \ce{&->O- + H2}\\
    \ce{&->OH- + H}.
\end{align}
In pure \ce{H2O} and mixed \ce{H2O}:\ce{CO2} ices, irradiated by heavy ions (Ni), \citet{Pilling2010b} found that hydrogen peroxide, \ce{H2O2}, can be formed following destruction of water
\begin{equation}
    \ce{H2O \leadsto H2O^{*} -> OH + H},
\end{equation}
followed by the radical-radical reaction
\begin{equation}
    \ce{OH + OH -> H2O2}.
\end{equation}
They also measured carbonic acid, \ce{H2CO3}, forming in their mixed ices, which was explained with the reaction pathway involving destruction of water, and electron attachment of either \ce{CO2} or \ce{OH},
\begin{equation}
    \ce{H2O \leadsto OH + H+ + e-},
\end{equation}
\begin{align}
    \ce{OH + e- &-> OH- }, \\
    \ce{CO2 + e- &-> CO2-}
\end{align}
with a reaction through bicarbonate \ce{HCO3-},
\begin{align}
    \ce{CO2- + OH &-> HCO3-} \\
    \ce{CO2 + OH- &-> HCO3-},
\end{align}
followed by reactions with the previously generated \ce{H+},
\begin{equation}
    \ce{HCO3- + H+ -> H2CO3}.
\end{equation}

Pure methanol ice has also been intensively studied \citep{Boyer2014, Sullivan2016, Basalgete2021a, Mejia2022, Ivlev2023, Mejia2024}. \citet{Boyer2014} found that DEA in pure methanol ice leads to a number of anions, in particular \ce{H-}, \ce{O-}, \ce{OH-}, \ce{CH-}, \ce{CH2-}, \ce{CH3-}, and \ce{CH3O-}. The production of \ce{O-} follows from
\begin{align}
    \ce{CH3OH \leadsto CH3OH- &-> O- + CH4} \\
    \ce{& -> O- + CH3 + H} \\
    \ce{& -> O- + CH2 + H2},
\end{align}
at different electron-impact energy thresholds (2.39 eV, 7.01 eV, and 7.24 eV, respectively). Their experiments indicated that \ce{H-}, \ce{O-} and \ce{CH3O-} are direct byproducts of \ce{CH3OH} DEA, while the others result from reactions with the primary byproducts and other reactants, e.g., for \ce{OH-},
\begin{equation}
    \ce{O- + C_nH_{2n+2} -> OH- + C_nH_{2n+1}},
\end{equation}
while for \ce{C_nH-}, the authors suggest that these can form via reactions with fast \ce{H-}, which are not efficiently measured and thus not well constrained. Further, electron irradiation can lead to radical formation \citep{Boyer2016, Arumainayagam2019}, e.g.,
\begin{align}
    \ce{CH3OH \leadsto CH3OH^* &-> CH3O + H \\}
    \ce{&-> CH2OH + H},
\end{align}
which can further and rapidly react with other species to form more complex molecules, such as
\begin{equation}
    \ce{CH3O + CH2OH -> CH3OCH2OH}
\end{equation}
\begin{equation}
    \ce{CH2OH + CH2OH -> HOCH2CH2OH}.
\end{equation}
Smaller fragments that are produced, such as \ce{CH3} and \ce{HCO}, can also react to form, e.g., acetaldehyde (\ce{CH3CHO}). 

The introduction of nitrogen into ice mixes enables a more interesting and complex chemistry due to the radicals that can form, such as \ce{NH} and \ce{NH2}. For instance ground-state \ce{NH2} can be formed through DEA and excited \ce{NH2} can form through electron and photon excitation \citep{Pilling2010a, Shulenberger2019}:
\begin{align}
    \ce{NH3 &\leadsto H- + NH2} \\
    \ce{NH3 &\leadsto NH2(\tilde{X}^2B_1) + H} \\
    \ce{NH3 + h\nu &-> NH2(\tilde{X}^2B_1) + H}.
\end{align}
It can also be formed through the ionization processes \citep{Shulenberger2019, Ciaravella2019}
\begin{equation}
    \ce{NH3 \leadsto NH3^+ + 2e- -> NH2 + H+ + 2e-},
\end{equation}
or
\begin{align}
    \ce{NH3 &\leadsto NH3^+ + 2e-} \\
    \ce{NH3+ + NH3 &-> NH2 + NH4+},
\end{align}
which has the added effect of creating \ce{NH4+}. The above processes can also lead to ion products
\begin{equation}
    \ce{NH3 \leadsto NH3^* ->} 
        \begin{cases}
            \ce{NH2+ + H+ + 2e_A^-} & \\
            \ce{NH+ + H+ + H + 2e_A^-}, &
        \end{cases}
\end{equation}
where \ce{e_A^-} denotes ejected auger electrons. The anion molecule \ce{OCN-}, which has been observed in the ices in molecular clouds and cores with the James Webb Space Telescope \citep{McClure2023, Sturm2023, Rocha2024, Chen2024, Rocha2025, Tyagi2025}, can be formed via
\begin{align}
    \ce{NH2 + CO &-> HNCO + H} \\
    \ce{NH3 + HNCO &-> OCN- + NH4+}.
\end{align}
The availability of these radicals also enables the formation of more complex nitrogen-bearing prebiotic molecules such as formamide \citep{MunozCaro2019} (\ce{HCONH2})
\begin{align}
    \ce{NH2 + HCO &-> HCONH2} \\
    \ce{NH2 + CO &-> NH2CO} \\
    \ce{NH2CO + H &-> HCONH2},
\end{align}
along with the formation of urea (\ce{NH2CONH2}),
\begin{equation}
    \ce{NH2CO + NH2 -> NH2CONH2}.
\end{equation}

\begin{figure}
    \centering
    \includegraphics[width=\columnwidth]{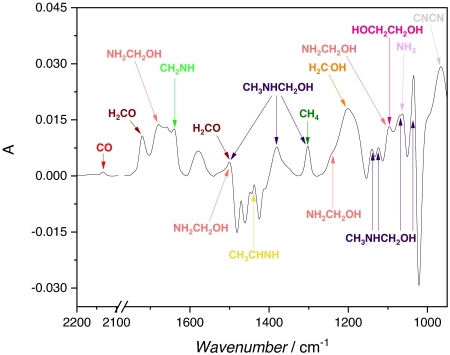}
    \caption{\label{fig:keresztes2024}Difference spectrum of electron irradiated (5 keV) \ce{CH3OH}:\ce{CH3NH2} mixed ice at 3 K. The spectrum is the difference between the spectrum recorded after electron irradiation and just after deposition. Reused with permission from \citet{Keresztes2024}.}
\end{figure}

Experiments of irradiation of ice with mixtures of water, carbon monoxide (or other carbon-bearing molecule such as methane), and ammonia (or other nitrogen-bearing molecules) \citep{Esmali2018, Jimenez-Escobar2018, Ciaravella2019, Vasconcelos2020, Jimenez-Escobar2022, Keresztes2024, Mate2025} have found that the resulting products can include complex organic molecules and prebiotics, including the possible production of glycine \citep{Esmali2018, Mate2025}. The exact formation pathways for these complex species have not been well elucidated, so they are not posited here. 

Figure \ref{fig:keresztes2024} shows recent experimental results from \citet{Keresztes2024}, where they irradiated pure \ce{CH3OH} and \ce{CH3NH2} ices and mixed \ce{CH3OH}:\ce{CH3NH2} ices with 5 keV electrons at the VIZSLA facility. They find a number of interesting results for the differences in the irradiation of the pure ices, with the formation of complex organics thought to originate through radical-radical combination. Crucially, they found that in the mixed ices, the molecules that were formed were different than the pure species, with molecules found in the pure ice irradiation experiments not found in the mixed ices. The difference spectrum shown in Figure \ref{fig:keresztes2024} demonstrates that the irradiation not only produces smaller fragments, but there is also the production of more complex organic and prebiotic molecules. The spectrum and their analysis is not able to measure all radicals produced due to how rapidly they interact in the ices.

There has been much interest in the ice chemistry of sulfur. How much, and in what form, sulfur is depleted onto grains is still an open question, in particular since \ce{H2S} has remained elusive in observations of ices \citep{Boogert2015, McClure2023}. Energetic radiation can play an important role in the processing of sulfur in ices, with the destruction of hydrogen sulfide and the increasing production of sulfanes (\ce{H2S_n}) and sulfur chains up to octasulfur (\ce{S8}). \citet{Shingledecker2020S} presented a theoretical study of sulfur in ices and found that the inclusion of radiolysis with nondiffusive reactions leads to the enhancement of \ce{OCS}, \ce{SO2} and \ce{S8} in ices. There has been a sizable amount of historical laboratory studies of sulfur-bearing ices \citep{Moore2007, Jimenez-Escobar2012, Mifsud2022, Carrascosa2024}, with two works of note most highlighting the role of energetic processes \citep{Mifsud2025, Herath2025}. \citet{Herath2025} irradiated \ce{H2S} ices with 5 keV electrons and found that ices are processed into sulfanes and sulfur chains through radiolysis followed by radical reactions. In particular, \ce{H2S} undergoes the radiolysis reactions
\begin{align}
    \ce{H2S & \leadsto HS + H}\\
    \ce{H2S & \leadsto S(^1D) + H2}.
\end{align}
Disulfane, \ce{H2S2}, is then formed through radical-radical reactions or via the electronically-excited sulfur,
\begin{align}
    \ce{HS + HS &-> H2S2} \\
    \ce{H2S + S(^1D) &-> H2S2 }. 
\end{align}
The buildup of complexity is then followed by either radiolysis of products, e.g.,
\begin{equation}
    \ce{H2S2 \leadsto H + HS2},
\end{equation}
followed by atomic sulfur injections \ce{S + H2S_X -> H2S_{(X+1)}} or reactions with radicals, e.g., \ce{HS_X + HS_Y -> H2S_{(X+Y)}}. 

\citet{Mifsud2025} irradiated an ice composed of \ce{O2}, CO, \ce{CO2}, or \ce{H2O} pure ices on top of allotropic sulfur with 1 MeV He+ ions. Their experiments found the formation primarily of species such as \ce{SO2}, \ce{CS2} and \ce{OCS} but no \ce{H2S}, in qualitative agreement with recent JWST observations of ices \citep{McClure2023}. The experiments agreed with previous works that \ce{SO2} forms via direct energetic processes while \ce{OCS} form via non-energetic processes with radicals, which are themselves enhanced via radiolysis.

X-ray irradiation of ices with multiple layers was also found to increase the diffusivity in the ice, enabling molecules from the bulk to react more readily with those in lower layers or the substrate \citep{Jimenez-Escobar2022}. The formation of complex prebiotics, such as glycine, due to non-energetic atom addition or radical-radical ices has also been seen in laboratory experiments \citep{Krasnokutski2020, Ioppolo2021}. The enhanced presence of radicals due to irradiation may increase the efficiency of this mechanism. While there has been a growing number of investigations in the laboratory work of high-energy ice astrochemistry, numerical improvements have primarily focused on cosmic-ray-induced desorption processes, with the improved radiolysis methodology of \citet{Shingledecker2018a} not yet being widely adopted, nor the inclusion of X-ray radiation processes. 

Finally, there has been a growing interest in the role of highly energetic processes in the development of homochiral chemistry. The homochirality of life has been well established since discovered by Louis Pasteur. Recently, there was the first detection of a chiral molecule, in this case propylene oxide (\ce{CH3CHCH2O}) in space \citep{McGuire2016}. It has been proposed that spin-polarized electrons can produce an asymmetry in the chirality \citep{Rosenberg2008, Rosenberg2019}. Laboratory simulations have shown that both X-rays and energetic particle irradiation of magnetic substrates can produce spin-polarized electrons \citep{Pfandzelter2003, Rosenberg2015}. Finally, it was recently proposed by \citet{Hoang2025} that cosmic-ray irradiation of magnetically-aligned grains can induced spin-polarized electrons and potentially provide a means to induce a chiral asymmetry in star- and planet-forming gas. However, the work did not include a chemical model, and so future work must be done to investigate the role of induced spin-polarized electrons via high-energy radiation on the chirality of organic and prebotic molecules.

\subsection{Demonstration through UCLCHEM models}
\begin{figure*}
    \centering
    \includegraphics[width=\textwidth]{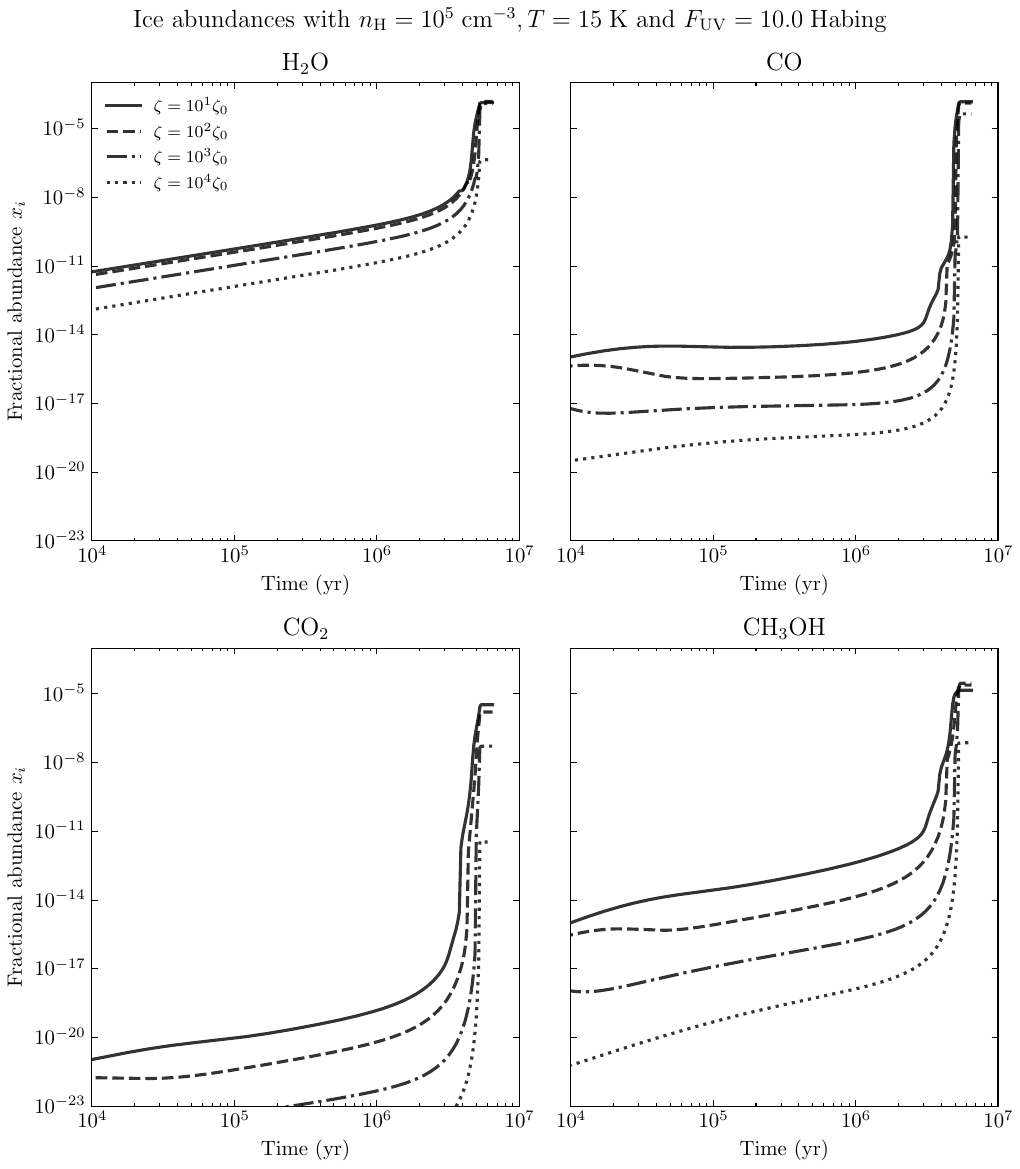}
    \caption{\label{fig:UCLCHEM} Abundances of \ce{H2O} (top left), \ce{CO} (top right), \ce{CO2} (bottom left), and \ce{CH3OH} (bottom right), in the ice as a function of time for a one-zone chemical model using the {\sc UCLCHEM} code. The different lines denote the annotated ionization rate. }
\end{figure*}
In this section, we present models which utilize the public three-phase gas-grain code {\sc UCLCHEM}\footnote{\url{https://uclchem.github.io/}} \citep{Holdship2017}. {\sc UCLCHEM} solves for the abundances of gas-phase molecules and the abundances of ice species, including both the surface and the mantle species, with a wide range of included chemical processes. Figure \ref{fig:UCLCHEM} shows an isothermal {\sc UCLCHEM} one-zone model of a dense gas region. The model follows the free-fall collapse of a gas parcel \citep{Rawlings1992} for 5 Myr with an initial gas density of $100$ cm$^{-3}$ and initial extinction of $A_V = 2$ mag, fixed gas and dust temperature of $15$ Kelvins, and an external FUV field of 10 Habing \citep{Habing1968}, with a final extinction of $A_V = 100$ mag. The simulations come from \citet{Dutkowska2025}, with the parameter space in their Table 2. The models are run for values $10 - 10^4$ times the fiducial ionization rate, $\zeta_0 = 1.3\times^{-17}$ s$^{-1}$. These models do not include the bulk radiolysis chemistry, but include cosmic-ray desorption processes. The models show that ice freezeout and chemistry are rapid. As the ionization increases, in all cases, the amount of each molecule in the ice decreases. Water is less impacted due to how tightly bound it is to the surface. 

\section{Chemical modeling including high-energy radiation - state of the art}\label{sec:chemmodel}
Chemical models are essential tools that allow us to predict molecular abundances as a function of a large range of physical conditions. Many physical and chemical input parameters influence the outcome of a chemical model, including, of course, the amount of high-energy radiation (cosmic rays or X-rays) that is available to the simulated astronomical system. 

While X-ray chemistry is seldom included in chemical codes, they ubiquitously include cosmic ray-induced chemistry, and in most reaction networks, the default rate of cosmic ray ionization of hydrogen is set to the typical value for the Milky Way Galaxy; this rate can then be enhanced or reduced in all chemical models. The importance of cosmic rays can not be overstated, as described in the sections above: they initiate the chemistry in quiescent conditions in places deep inside the molecular clouds where UV radiation does not penetrate.  We have seen in Section \ref{sec:gasphase} that cosmic rays have a wide range of effects: they produce atomic hydrogen via dissociation of \ce{H2}, they are the dominant source of ionisation, they provide heat and charge to the dust grains, and they regulate the degree of coupling between the gas and the magnetic field and hence have a key role to play in the collapse timescale that leads to protostars \citep{Wurster2018, Grassi2019, Tritsis2022}. Hence, it is of particular importance to accurately treat them in chemical models. 

In the past, all chemical models simply assumed a fixed initial cosmic ray ionization flux that did not vary with time nor depth into the cloud. However, as described in Section \ref{sec:gasphase}, it is now clear that the cosmic ray ionization rate is a function of column density and such dependency needs to be included in chemical models. It has in fact been shown \citep{Rimmer2012, O'Donoghue2022, Gaches2019, Gaches2022, Latrille2025} that, especially during the early collapse of a molecular cloud, the attenuation of the cosmic ray ionization deeper in the cloud has a destructive effect for some of the key species such as CS, NH$_3$, and solid CO$_2$. Perhaps more importantly, the chemical effects of the inclusion of the dependencies of cosmic ray flux in a model leads, in a non-linear way, to different chemistry depending on the physical conditions of the cloud, such as volume densities and gas temperatures. Several codes are now able to treat such dependency, such as {\sc UCLCHEM} \citep{O'Donoghue2022} and {\sc 3D-PDR} \citep{Gaches2019, Gaches2022}, using built-in functions, and {\sc Nautilus} \citep{Ruaud2016}\footnote{\url{https://forge.oasu.u-bordeaux.fr/LAB/astrochem-tools/pnautilus}} via the user providing a one-dimensional profile in the initial condition text file.

Cosmic rays not only affect the gas phase chemistry. They also interact with the grains, aiding chemical reactions and non-thermal desorption of the icy mantles.  This is particularly important in cold molecular clouds where diffusive chemistry may not be very efficient due to low thermal energies \citep{Ghesquiere2018}. In laboratory experiments, \citet{Ghesquiere2018} found that bulk chemistry may be regulated through structural changes in the ice that enable the transport of molecules in the bulk, especially radicals, rather than thermal bulk diffusion. Cosmic-ray-induced desorption can be quite complex, especially due to the interplay between grain heating frequencies and cosmic ray fluxes. Several studies have now shown that desorption due to heating of the dust by cosmic rays may be even more important than thermal desorption \citep{Shingledecker2018b, Kalvans2019}. Besides aiding desorption, cosmic rays irradiation of ices can also lead to the formation of complex organic molecules (e.g. \citet{Modica2010}), enhanced ionization on the dust grains \citep{Ivlev2015b} or an enhancement of excited species \citep{Shingledecker2018a}. Chemical models vary in the degree to which they include such effects, although public codes have so far focused on non-thermal desorption processes. The radiolysis model of \citet{Shingledecker2018a} was applied using the {\sc Nautilus} code \citep{Shingledecker2018b, Shingledecker2019} and has also been applied using the {\sc MONACO} \citep{Vasyunin2017} code for \ce{O2} and \ce{H2O} ice \citep{Shingledecker2019}.

The effects of X-rays are seldom included in gas-grain chemical models. Gas phase models which include the effects of X-rays have however, existed for decades \citep{Spaans2005, Meijerink2005, Priestley2017, Gunasekera2025}. The main effect of X-rays is to heat the cloud deeper than UV photons and this leads to excitation as well as, to a lesser extent, chemical effects. Hence, fine-structure line ratios differ between PDRs and XDRs, and the CO higher J transitions are more populated due to the X rays irradiation. Chemical models that treat the impact of X-rays the same as cosmic rays will be neglecting much of the induced and complicated line cooling that is created by the preferential ionization (and Auger ionization) of X-rays compared to cosmic rays. Of interest, especially for extragalactic studies, may be the effects of X-rays on ratios such as HCN/\ce{HCO+}, which can be less than 1 in X-ray-dominated regions. 

It is worth noting here that while most chemical models make use of publicly available and maintained gas phase chemical reactions networks such as UMIST \citep{Millar2024}\footnote{\url{https://umistdatabase.uk/}} and KIDA \citep{Wakelam2024}\footnote{\url{https://kida.astrochem-tools.org/}},  grain surface reaction networks are generally created for the purpose of specific studies by the users and as such it is not possible to provide a generalised summary of what is at the moment included in chemical models with regard to cosmic rays effects on the grains.  Nevertheless, we refer the reader to a recent modelling effort by \citet{Shingledecker2018a} where they account for the ice radiolysis that results from the interaction of ionizing radiation and the surface of the grains.  

Recently, Pilling and collaborators have introduced the PROCODA code, which has been utilized to infer reaction rates for ices undergoing radiolysis from energetic particle or X-ray irradiation \citep{Pilling2022}. The code enables a statistical fitting of reaction rates from experimental ice data. While the constrained reaction rates have not seen as much usage, in part because most astrochemistry gas-grain codes calculate the ice-phase reaction rates from fundamental data such as binding or diffusion energies, the results allow unique insights into the chemical pathways induced by high-energy radiation. The code can fit many hundreds of reaction rates, although since the laboratory data typically does not have that many temporal and species data points, the results have so far relied on assumed constraints on the chemistry and on interpolating the experimental data onto finer temporal spacings to reduce the degeneracy of the fits.

\section{Laboratories of interest for high-energy astrochemistry in space}\label{sec:casestudies}
\subsection{Active Galactic Nuclei and Starburst regions}
Cosmic rays and/or X-ray fluxes are found to be enhanced, compared to those found in the Milky Way, in both Active Galactic Nuclei and Starburst galaxies. Hence, we shall use these two types of objects as examples of extra-galactic laboratories of high-energy astrochemistry.

Active Galactic Nuclei (AGN), bright and compact objects at the center of galaxies, are well-known sources of cosmic rays and X-rays. This high-energy radiation affects the surrounding medium by heating and ionizing it and is therefore an essential component of the feedback mechanisms that act in galaxies.  The effects of such feedback can be indirectly studied by exploring the cosmic ray and X-ray-driven chemistry that occurs in the gas and on the dust in the vicinity of an AGN. Single dish molecular observations of nearby AGNs such as NGC 1068 already show that the chemistry is strongly influenced by cosmic rays (e.g., \citet{Aladro2013}), although \citet{Esposito2022} shows that such effects die out at scale $\geq 250$ pc. Closer to the AGN (inside the circumnuclear disk) interferometric observations with ALMA confirm that the temperatures near the AGN are high compared to the starburst ring in NGC 1068 and that, more importantly, the \ce{HCO+} emission can only be matched by an enhanced (by at least a factor of 100) cosmic ray ionization rate compared to that measured in our Milky Way \citep{Viti2014}.
We recall, however, that disentangling the chemical effects due to cosmic rays from those due to X-rays on the dense molecular gas in the vicinity of an AGN is not trivial, and multiple transitions of the molecules CO, HCN, and \ce{HCO+} are needed. In particular, an HCN/\ce{HCO+} ratio above 1 has often being invoked as a robust indicator of AGN activity \citep{Kohno2008}: however this is not always the case and a correlation with an enhancement of the HCN/\ce{HCO+} (1-0) line ratio with AGN failed the statistical test when applied to a large sample \citep{Privon2020}. In fact, it has now been shown, albeit for a limited number of sources, that whether this ratio is a robust tracer of AGN activity or not depends strongly on the spatial resolution of the observations as well as the J transitions used \citep{Butterworth2025}. 

\begin{figure}
    \centering
    \includegraphics[width=\columnwidth]{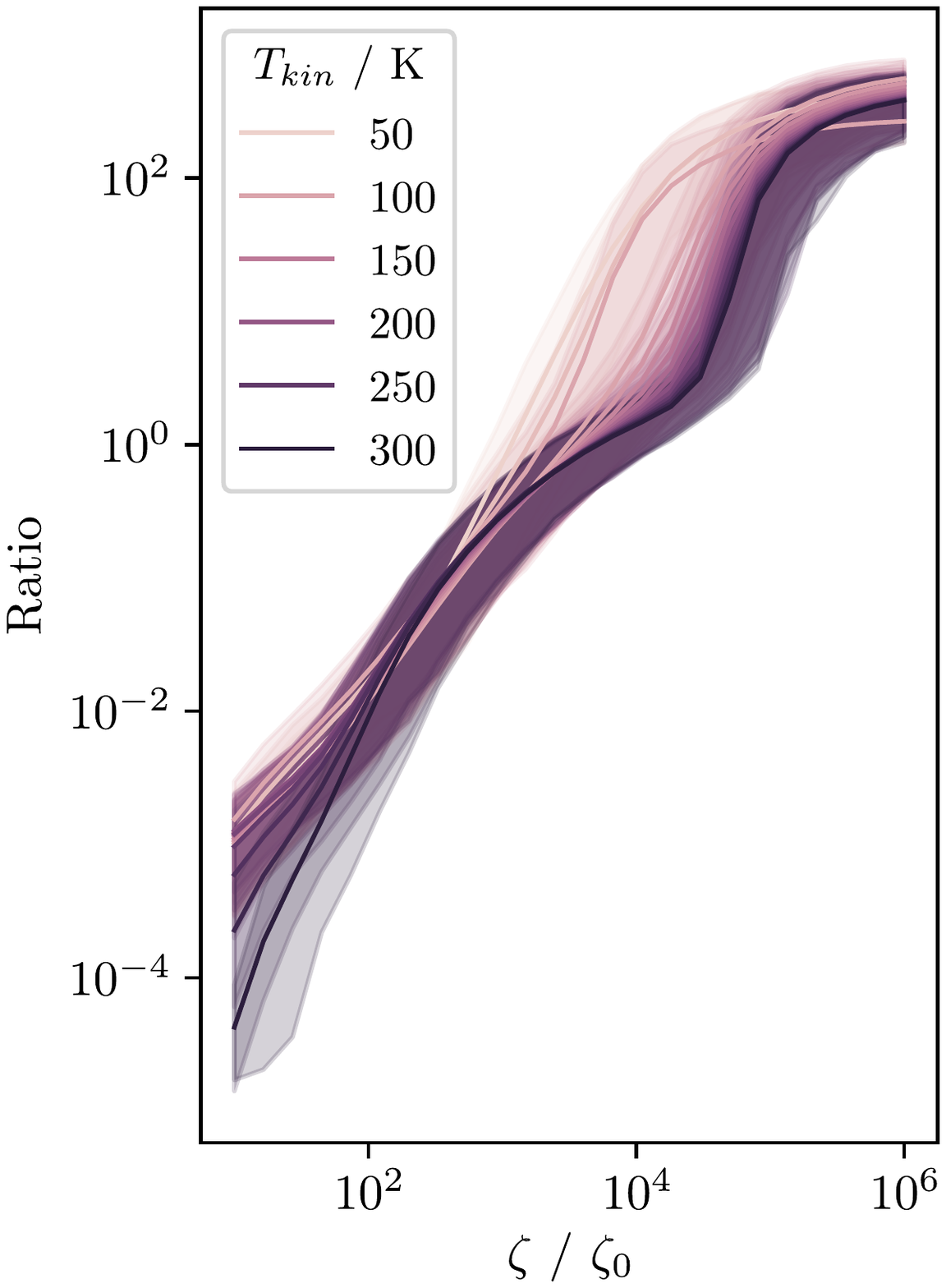}
    \caption{\label{fig:holdship2022} The ratio of \ce{H3O+}/\ce{SO} as a function of the cosmic-ray ionization rate in units of the ``fiducial'' ionization rate, $\zeta_0 = 1.3\times 10^{-17}$ s$^{-1}$ for different gas temperatures. Adapted with permission from \citet{Holdship2022}.}
\end{figure}

Starburst galaxies, sites of extreme (massive) star formation as compared to the Milky Way, are the perfect laboratories to study the process of star formation in very high-density, high-temperature environments. One of the many processes associated with massive star formation is strong ionization by cosmic rays from supernova remnants. In fact, clear chemical effects of an enhanced cosmic ray flux have now been shown for the most nearby galaxies such as NGC 253: the ALMA Large Program ALCHEMI \citep{Martin2021} delivered the most complete extragalactic molecular inventory in the central molecular zone of this galaxy at a spatial resolution comparable to Giant Molecular Clouds and studies of individual sets of molecules from this survey have consistently revealed a cosmic ray ionization rate which is $\sim$ 10$^{-14}$ s$^{-1}$ within 100 pc from the nucleus. For example, while the HCN/HNC ratio decreases with an increase of cosmic ray ionization rate \citep{Behrens2024},  the H$_3$O$^+$/SO ratio increases, and in fact shows a surprisingly strong correlation with cosmic ray ionization rate \citep{Holdship2022}. Figure \ref{fig:holdship2022} shows this ratio as a function of gas temperature and $\zeta$, averaged over models with a range of densities between $10^4$ and $10^7$ cm$^{-3}$. The ratio is a sensitive probe of $\zeta$ because of how the ionization rate acts on \ce{H3O+} and SO inversely: \ce{H3O+} is primarily produced via ionization-driven processes (see above), while SO is destroyed primarily by atomic ions such as \ce{C+} and \ce{H+}, both of which are dominantly produced via cosmic-ray driven processes in dense gas. Due to this, the ratio becomes highly sensitive to $\zeta$.

\subsection{Resolved molecular clouds}
\begin{figure}
    \centering
    \includegraphics[width=\columnwidth]{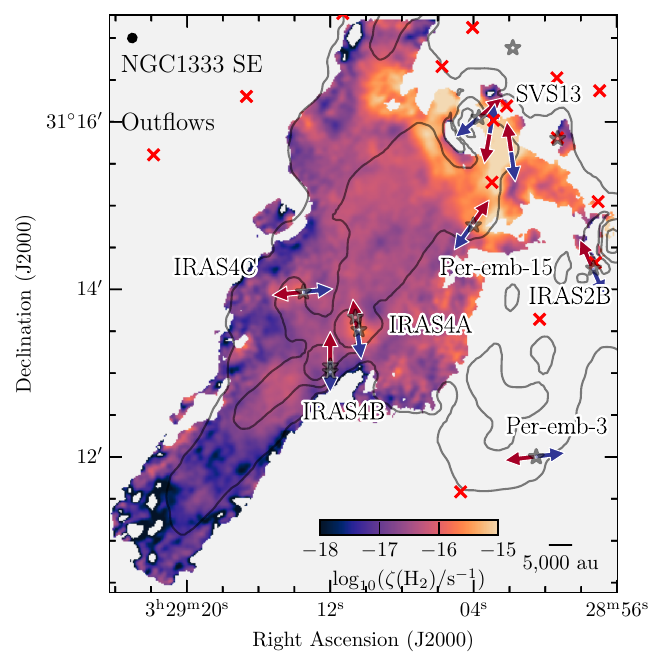}
    \caption{\label{fig:pineda2024}Estimated ionization rate throughout the NGC 1333 star-forming region. NGC 1333 is located in the Perseus cloud at a distance of 300 pc. Overlaying contours denote the \ce{H2} column density. The stars mark the location of young stellar objects (YSOs). Adapted with permission from \citet{Pineda2024}.} 
\end{figure}
Molecular clouds, many of which host star-forming regions, can serve as interesting probes of high-energy astrochemical processes, in particular massive star-forming regions. Recently, there has been significant progress in producing resolved maps of the cosmic-ray ionization rate \citep{Sabatini2023, Pineda2024, Socci2024} that also show correlations of enhanced ionization rates towards nearby sites of star formation. Figure \ref{fig:pineda2024} shows the estimated ionization rate map derived from a combination of \ce{H^{13}CO+}, \ce{DCO+}, and \ce{C^{18}O}, for the nearby star-forming region NGC 1333. There are nodes of enhanced ionization rate towards locations of embedded star formation, suggesting there are embedded sources of energetic radiation. 

\begin{figure}
    \centering
    \includegraphics[width=\columnwidth]{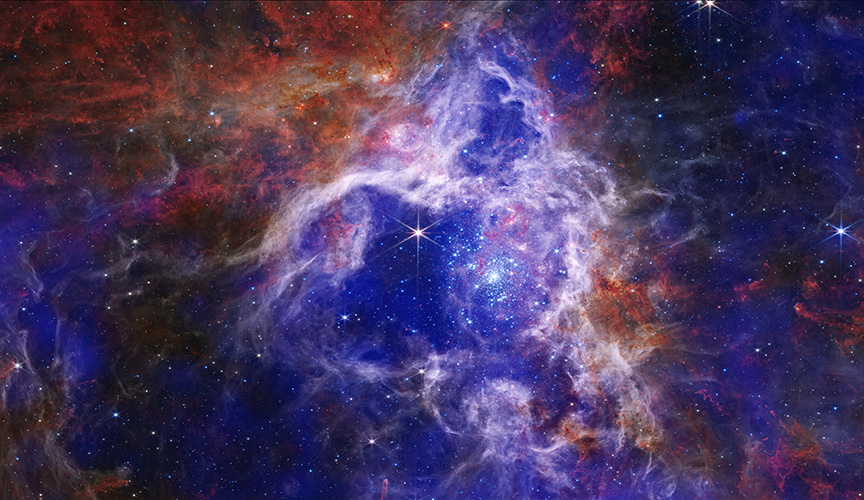}
    \caption{\label{fig:30dor}30 Doradus, also known as the Tarantula Nebula, located in the Large Magellanic Cloud, as viewed by the Chandra X-ray Observatory (royal blue and purple) and the James Webb Space Telescope (red, orange, green, and light blue). X-ray: NASA/CXC/Penn State Univ./L. Townsley et al.; IR: NASA/ESA/CSA/STScI/JWST ERO Production Team}
\end{figure}

There have been a number of X-ray observations towards star formation, but two important X-ray surveys to highlight are of massive star-forming regions:\ ``The Massive Star-forming Regions Omnibus X-ray Catalog'' \citep{Townsley2014, Townsley2018, Townsley2019} and ``Massive Young Star-Forming Complex Study in Infrared and X-Ray (MYStIX)'' \citep{Feigelson2013}. These surveys show that massive star-forming regions are immersed in diffuse X-ray radiation. Molecular clouds in local galaxies are also resolved now across the electromagnetic spectrum: Figure \ref{fig:30dor} shows the region 30 Doradus, or the Tarantula Nebula, in X-ray and infrared emission, highlighting how the diffuse X-ray emission fills the cavity, irradiating the walls of the cavity with intense X-ray emission. Many molecular clouds are also seen being impacted by supernova remnants \citep{Ceccarelli2011, Schuppan2012, Vaupre2014, Sofue2021, Zhou2022, Cosentino2022, Indriolo2023, Cosentino2025}. Finally, molecular clouds in the galactic center are also known to be experiencing enhanced ionization rates \citep{Oka2005, Geballe2010, Zeng2018, Rivilla2018, Oka2019}

\begin{figure}
    \centering
    \includegraphics[trim={1.5cm 2cm 2.5cm 3.5cm}, clip, width=\columnwidth]{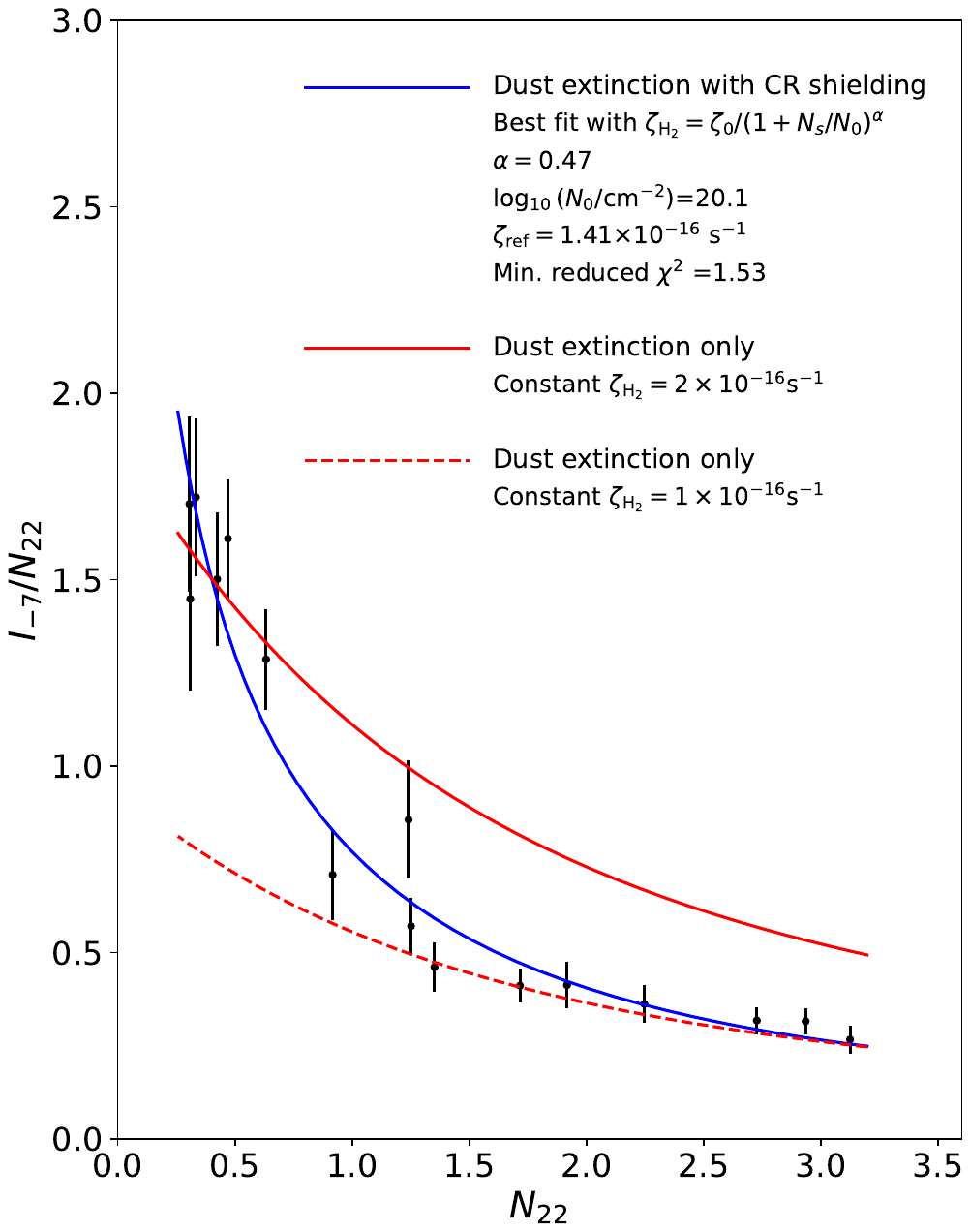}
    \caption{\label{fig:neufeld2025}Ratio of the intensity of the 1-0 O(2) \ce{H2} line, $I_{-7}$ in units of $10^{-7}$ erg s$^{-1}$ cm$^{-2}$ s$^{-1}$, and total \ce{H2} column density, $N_{22}$ in units of $10^{22}$ cm$^{-2}$ versus $N_{22}$. The lines show different cosmic-ray models as annotated in the legend (see text). Adapted with permission from \citet{Neufeld2025}, provided by D. Neufeld.}
\end{figure}

The James Webb Space Telescope has opened up a new avenue of constraining the ionization rate: recently, \citet{Bialy2025} robustly detected the cosmic-ray-induced near-infrared \ce{H2} emission in the Bok globule Barnard 68 (B68). The detection was strong enough to even begin to constrain the gradient within the cloud \citep{Neufeld2025}. Figure \ref{fig:neufeld2025} shows the main result of \citet{Neufeld2025}, namely that the observed reduction in the 1-0 O(2) line with column density into B68 can only be explained with cosmic ray attenuation. The constrained attenuation profile was shown not to be consistent with an attenuation model including only energy losses, hinting that transport effects with magnetic fields, such as shielding or diffusion, may be necessary. This new method, if able to be used for many more objects, will provide tight constraints on the ionization rate and particle transport in nearby clouds.

\subsection{Protoplanetary Disks}
Protoplanetary disks are an evolved stage of protostellar disks, when the surrounding cores have largely dissipated, exposing the disks. Low-mass protostars in this stage, T-Tauri stars, have magnetic fields that create magnetospheres around them, dubbed ``T-Tauriosphere'', that exclude galactic cosmic rays \citep{Cleeves2013, Cleeves2015}. Their magnetic fields interact with the surrounding disk, leading to reconnection events and flaring \citep{Shu1997, Feigelson1999, Feigelson2007, Getman2021, Getman2022}, producing high-energy radiation. Protostars also exhibit bursts of accretion, enhancing their X-ray and energetic particle flux \citep{Lee1998, Rab2017, Brunn2023, Brunn2024}.

The variable high-energy radiation leads to fluctuations in the abundances of light ions and specific neutrals \citep{Waggoner2022}. The steady X-ray radiation and particle radiation impact the abundance of key ions and heats the gas \citep{Aresu2011, Rab2017, Rodgers-Lee2020, Washinoue2024}.Protoplanetary disk chemical models have maintained sustained advances in X-ray chemistry modeling, including variability \citep{Meijerink2012, Rab2018, Waggoner2019, Sellek2024}. The impact of protostellar energetic particles, powered by accretion or reconnection events, on disks has also been modelled recently \citep{Rab2017, Offner2019, Rodgers-Lee2020, Brunn2023}. Therefore, these regions make ideal laboratories to investigate short-time variable high-energy fluxes.

\section{Summary Points}
In this review, we have summarized the gas- and ice-phase chemistry induced by high-energy radiation. Here, we have defined this as ionizing radiation capable of producing a significant cascade of secondary electrons. Below, we summarize the main takeaway points:
\begin{enumerate}
    \item There are a plethora of sources which emit high-energy radiation, including both X-rays, with energies exceeding 500 eV, and energetic particles. Some of the sources, in particular high-mass stars -- both during their main sequence and during supernova -- and protostars, are tightly associated with nearby natal molecular gas.
    \item Ionizing radiation helps drive a diverse gas-phase chemistry after initiating ion-neutral reaction pathways, which can proceed rapidly without an energy barrier. These pathways are initiated primarily through the ionization of \ce{H} and \ce{H2}, with the latter forming the crucial trihydrogen cation, \ce{H3+}, and \ce{He+}, which is an important destroyer of molecular bonds. The initiation of deuteration is largely brought on by high-energy irradiation.
    \item Cosmic rays efficiently heat dense gas, and for ionization rates exceeding $10^{-16}$ s$^{-1}$, they dominate the heating budget for dense gas in average Milky Way environments. X-ray radiation heating is very efficient in dense, molecular gas due to the impact of Auger ionization, leading to much of the initial X-ray photon energy being converted into thermal energy. While hard X-ray photons dominantly ionize metals, with a significant impact of Auger processes, energetic particles dominantly singly ionize molecular hydrogen and helium, with heavier species being a smaller correction factor.
    \item High-energy ice chemistry, radiolysis, is driven primarily by low-energy secondary electrons which are induced following primary ionization of species both in the ice and the mineral substrate. In the ice bulk, electrons with energies below 20 eV drive a complex chemistry following the excitation of molecules and production of radicals. The production of cations from the primary ionizations and high-energy secondary electrons, and anions from dissociative electron attachment, drives a rapid ion-neutral chemistry. Radiolysis creates an elevated abundance of radicals, enabling a rapid build-up of molecular complexity, leading to the formation of complex organics and prebiotics, including the possibility of glycine. 
    \item Ionization and heating in the ices lead to sputtering and desorption of molecules from the surface. The sputtering is dominated by heavy ions due to the scaling of sputtering yields with $Z^4$. Desorption occurs both by cosmic-ray and X-ray-induced thermal desorption, because of the thermal heating, and via non-thermal processes resulting from the production of cations and the resulting exothermic reactions.
    \item Chemical models now include a variety of high-energy radiation processes, although there is still significant progress to be made. Much of the focus has been on cosmic ray chemistry, with a substantial paucity in improvements in X-ray chemical models in the past decade, with the exception of protoplanetary disk chemistry. On cloud scales, there has been some work incorporating X-rays into magnetohydrodynamic simulations with chemistry. In chemical codes, X-rays are typically included by treating the X-ray ionization rate the same as the CRIR. Astrochemical codes are now including cosmic ray gradients due to the energy loss functions, in large part due to the availability of polynomial approximate functions for $\zeta(N)$. Gas-grain models now include non-thermal desorption and sputtering processes for many molecules of importance.
    \item Observations, in particular with a combination of infrared observatories, such as the James Webb Space Telescope, and radio observatories such as the Atacama Large Millimeter Array and the Northern Extended Millimetre Array, and proposed telescopes like the Atacama Large Aperture Submillimeter Telescope, can now help detail the complex gas and ice chemistry in key laboratories of high-energy astrochemistry. Such joint observational programs will be key in the coming decade to advancing our understanding of high-energy astrochemical processes.
 \end{enumerate}
\section{Future Issues}
We highlight here key issues that need to be addressed in high-energy astrochemistry
\begin{enumerate}
    \item It has become increasingly clear \citep{Gaches2022, Latrille2025} that the inclusion of cosmic-ray gradients in chemical models can noticeably impact the modeled abundances and thus predicted molecular line fluxes. While the inclusion of more accurate cosmic-ray physics is growing \citep{Grassi2019, Gaches2019, Gaches2022, O'Donoghue2022, Redaelli2025}, there is still a substantial paucity of X-ray chemical models for cloud environments, with currently the only public code with chemistry and an accurate treatment of X-ray radiation processes being the {\sc Cloudy} spectral synthesis code \citep{Gunasekera2025}.
    \item X-ray-induced ionization of gas-phase molecules is primarily driven by secondary electrons. As of yet, there have been no chemical models including the electron-impact ionization of all observationally important species. Historically, this was constrained by a lack of cross-section data, but the new Astrochemistry Low-energy electron Cross-Section database \citep{Gaches2024} now includes a growing number of molecules ($\approx$200 in the first release). The other main constraint is the lack of (dissociative) recombination rates for many cations, inhibiting the proper inclusion of their ionization chemistry.
    \item While there has been some initial atomic-scale theoretical calculations of CR-induced ice chemistry \citep{Mainitz2016, Mainitz2017}, these used prescribed energy deposition tracks. A quantum mechanical atomicist theoretical view of high-energy ice interactions is still lacking, as also noted in \citet{Ceccarelli2023}. Therefore, a complete understanding of the chemical pathways is still not well constrained for radiolysis in ices for many molecules.
    \item There is a substantial lack of inclusion of radiolysis chemistry in gas-grain models, with the public codes including CR-induced thermal desorption, non-thermal desorption, and sputtering. However, there has not been a widespread adoption of the bulk radiolysis chemistry, such as that proposed by \citet{Shingledecker2018a}. The inclusion of these processes may be essential to model the ice chemistry of complex organic chemistry and the chemistry of sulfur-bearing molecules in the ice.
    \item There is a stark lack of chemical codes that include a full treatment of X-ray processes. While the code {\sc Cloudy} includes X-ray radiation transfer and a model of chemistry, it is not meant to be a primary astrochemical model. Development of public chemistry codes with an accurate treatment of X-ray radiation heating and chemistry is necessary to model regions, such as AGN, in which X-rays are thought to be important, rather than equating CR and X-ray processes. The inclusion of X-ray-induced radiolysis is also substantially lacking in gas-grain chemical models.
    \item The direct excitation of the rovibrational and electronic states of \ce{H2} by cosmic-ray protons \citep{Padovani2025} and secondary electrons \citep{Gredel1995, Bialy2025} may provide an alternative route to excited state chemistry, especially in high-ionization environments. However, the excitation of \ce{H2} by non-thermal radiation has yet to be included in any astrochemical model. These, and other non-thermal excitations, may prove to open up chemical pathways in energetic environments such as near active galactic nuclei.
\end{enumerate}

\begin{acknowledgments}
The authors acknowledge the outstanding contributions that Prof Eric Herbst has made, and is still making, towards the studies of high energy astrochemistry: his invaluable insights, published in many papers, as well as shared with us during our many conversations over the years have been a true source of knowledge!  We thank Marco Padovani for providing calculated data for the ranges of protons, electrons, and photons in molecular gas and for interesting discussions during writing. We thank Emmanuel Dartois, Tarczay György, David Neufeld, and Jaime Pineda for permission to reuse figures. BALG is supported by the German Research Foundation (DFG) in the form of an Emmy Noether Research Group - DFG project \#542802847 (GA 3170/3-1). SV is funded by the European Research Council (ERC) Advanced Grant MOPPEX 833460.vii.

\noindent The authors declare no competing financial interests.
\end{acknowledgments}

\bibliographystyle{aasjournal}
\bibliography{heac, manual}

\end{document}